\pdfoutput=1

\documentclass[11pt]{article}

\usepackage[preprint]{acl}

\usepackage{times}
\usepackage{latexsym}

\usepackage[T1]{fontenc}

\usepackage[utf8]{inputenc}

\usepackage{microtype}

\usepackage{inconsolata}

\usepackage{graphicx}
\usepackage{tcolorbox}
\usepackage{enumitem}
\usepackage{tabularx}
\usepackage{ragged2e}
\usepackage{booktabs}
\usepackage{multirow}
\usepackage{float}
\usepackage{listings}
\usepackage{xcolor}
\usepackage{caption}
\usepackage{enumitem}
\usepackage{subcaption}
\usepackage[normalem]{ulem}
\usepackage[table,xcdraw]{xcolor}
\usepackage{minted}
\useunder{\uline}{\ul}{}

\lstdefinestyle{promptstyle}{
    basicstyle=\ttfamily\small, 
    breaklines=true,            
    frame=single,               
    columns=fullflexible,       
    keepspaces=true,            
    backgroundcolor=\color{white}, 
    literate={*}{{\char42}}1    
             {-}{{\char45}}1
}

\definecolor{promptbgcolor}{RGB}{245,245,245} 
\definecolor{promptframecolor}{RGB}{180,180,180} 

\title{IRB: Automated Generation of Robust Factuality Benchmarks}

\author{
    Lam Thanh Do\textsuperscript{1}\thanks{Equal contribution},
    Bhagyashree Taleka\textsuperscript{2}\footnotemark[1],
    Hozaifa Ammar Bhutta\textsuperscript{1}, \\
    \textbf{Vikram Sharma Mailthody\textsuperscript{2},
    Kevin Chen-Chuan Chang\textsuperscript{1},
    Wen-mei Hwu\textsuperscript{2}} \\ 
    \textsuperscript{1}University of Illinois Urbana-Champaign \quad
    \textsuperscript{2}NVIDIA \\
    \texttt{\{lamdo, hbhutta, kcchang\}@illinois.edu} \\
    \texttt{\{btaleka, vmailthody, whwu\}@nvidia.com}
}

\definecolor{lamdocolor}{RGB}{220,20,60}

\begin{document}
\maketitle
\begin{abstract}

Static benchmarks for RAG systems often suffer from rapid saturation and require significant manual effort to maintain robustness. To address this, we present IRB, a framework for automatically generating benchmarks to evaluate the factuality of RAG systems. IRB employs a structured generation pipeline utilizing \textit{factual scaffold} and \textit{algorithmic scaffold}. We utilize IRB to construct a benchmark and evaluate frontier LLMs and retrievers. Our results demonstrate that IRB poses a significant challenge for frontier LLMs in the closed-book setting. Furthermore, our evaluation suggests that reasoning LLMs are more reliable, and that improving the retrieval component may yield more cost-effective gains in RAG system correctness than scaling the generator\footnote{\url{https://github.com/Hozaifa-Bhutta/IRB}}.

\end{abstract}

\section{Introduction}

Large Language Models (LLMs) have demonstrated remarkable general-purpose capabilities, but they remain susceptible to generating factually inaccurate or hallucinated content \cite{rawte2023troubling, xu2024hallucination}. Retrieval-Augmented Generation (RAG) has emerged as a prominent technique to mitigate this issue by grounding model responses in external, verifiable knowledge sources \cite{lewis2020retrieval, shuster2021retrieval}. Consequently, developing rigorous benchmarks to evaluate RAG systems has received much attention from the research community.

This rapid evolution of LLMs, however, introduces a significant challenge to existing evaluation frameworks. Newer models, trained on vast, web-scale datasets, often memorize benchmark data \cite{sainz-etal-2023-nlp, golchin2023time, dong-etal-2024-generalization}. This data contamination compromises evaluation reliability, as it becomes difficult to disentangle whether a model is utilizing the retrieved context or recalling an answer from its parametric knowledge \cite{longpre-etal-2021-entity, chen2024benchmarking}. Consequently, to ensure that evaluation remains meaningful, there is a persistent need to develop new, challenging benchmarks that can effectively test the capabilities of SOTA models.

Such an endeavor, however, is traditionally labor-intensive and expensive, as most dataset creation requires significant human annotation. Some benchmarks are constructed entirely by human annotators \cite{kwiatkowski2019natural, vu2023freshllms, pham2025sealqa}, while others use humans-in-the-loop to filter, refine low-quality samples generated by LLMs \cite{li-etal-2023-halueval, chen2024benchmarking}. A cheaper and more scalable alternative is to rely entirely on LLMs for data curation. Yet, previous work exploring this fully-automated approaches \cite{zhu2024rageval, filice2025generating} overlook explicit mechanisms to control the generation process. This absence of guidance not only can result in unfaithful or low-quality samples, but also provide little to no control over the types of questions that are generated.

To address these limitations, we introduce \textbf{IRB}, a framework that automates benchmark creation through structured, scaffolded generation. Our pipeline mitigates the instability of purely neural generation by introducing two distinct forms of guidance. First, we utilize a \textit{factual scaffold} derived from human-written citing sentences in Wikipedia. This ensures that generated questions and answers are grounded in fact, as every sentence is supported by human-verified evidence. Furthermore, these citations preserve the complex information needs inherent in human composition, ranging from temporal sensitivity (requiring up-to-date evidence) to cross-lingual reasoning. In this design, the sentence provides the factual basis for QA generation, while the cited documents serve as the retrieval ground truth.

\begin{table*}[ht]
\centering
\resizebox{0.75\textwidth}{!}{

\small
\renewcommand{\arraystretch}{1.3} 
\begin{tabularx}{\linewidth}{@{} >{\RaggedRight}X c c l c >{\RaggedRight\arraybackslash}p{2.5cm} @{}}
\toprule
\multirow{2}{*}{\textbf{Question}} & \multicolumn{4}{c}{\textbf{Attributes}} & \multirow{2}{*}{\textbf{Answer}} \\
\cmidrule(lr){2-5}
 & \textbf{Language} & \textbf{Freshness} & \textbf{Topics} & \textbf{Hops} & \\
\midrule
Who is the winningest head coach of the Mississippi State Bulldogs women's soccer team from 2019 to 2024 who compiled an overall record of 62 wins, 35 losses, and 18 draws? 
& En & 2024 & Culture, Geo & 1 & James Armstrong \\
\addlinespace 

\addlinespace
What is the name of the largest gas supplier company in a specific region that experienced harsh winter conditions, heating shortages, and isolated cases of hypothermia during the 2025 Moldovan energy crisis--which occurred after the cessation of Russian gas supplies at the end of December 2024--and that cut off the gas supply roughly 9 months ago to several buildings of the Moldovan authorities located in the Security Zone and in the city of Bender (Tighina)?
& Ru & 2025 & Geo, Culture & 2 & Tiraspoltransgaz \\
\addlinespace

What geological formation located in Peru contains fossils of Sinoeocrinus lui? 
& En & 2024 & STEM & 1 & False-premise question \\

\bottomrule
\end{tabularx}

}
\caption{Sample questions and answers generated by IRB, using 29 September 2025 as the reference date.}
\label{tab:sample_questions}
\end{table*}

Second, we impose an \textit{algorithmic scaffold} to guide the synthesis of question-answer (QA) pairs. By transforming source text into an intermediate knowledge graph, we utilize graph-guided generation to enforce constraints on the LLM. This mechanism allows us to programmatically dictate output complexity, enabling the precise construction of multi-hop reasoning chains, lexical paraphrases, and false-premise traps.

\begin{figure*}[h]
\centering
\includegraphics[width=0.8\textwidth]{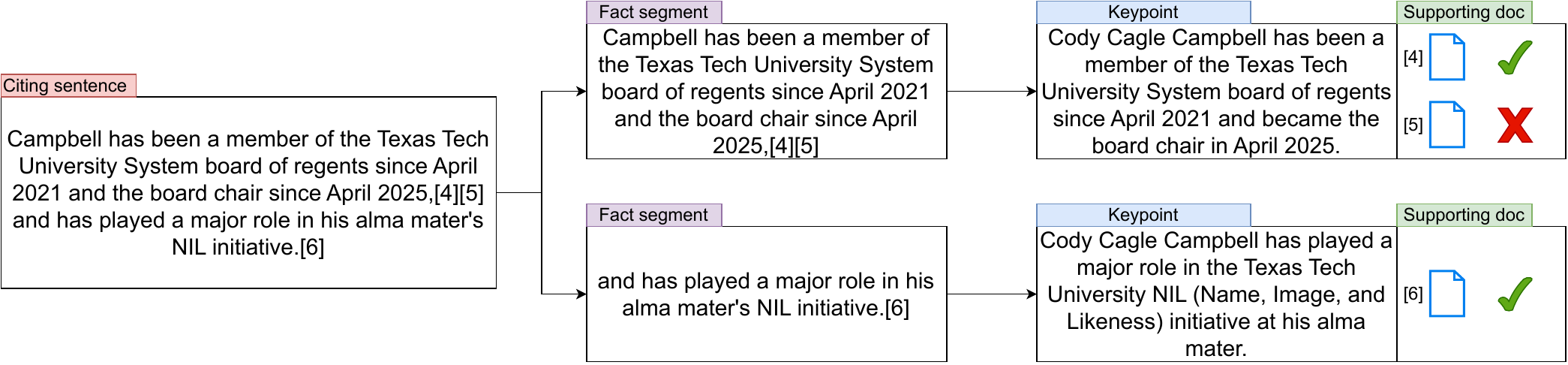}
\caption{Fact \& supporting documents extraction from a citing sentence. The resulting fact contains two keypoints, each with different supporting documents. Groundedness check is performed on each (keypoint, document) pair.}
\label{fig:IRBFactAndSupportingDoc}
\end{figure*}

Using the IRB framework, we generate IRB1K, to evaluate frontier LLMs and retrievers. Our evaluation reveals that IRB1K poses a significant challenge for frontier models in the closed-book setting. Additionally, while retrieval acts as an ``equalizer'' that aligns the correctness of various frontier LLMs, reasoning models prove to be more reliable; particularly in adversarial settings involving incorrect retrieval, false-premise questions and internal-external knowledge conflicts. Finally, our results suggest that given the high baseline capabilities of current frontier LLMs, larger and more cost-effective gains in RAG system correctness are best achieved by improving the retriever component.

We summarize our contributions as follows: 1) We propose IRB, an extensible framework for generating granular RAG evaluation benchmarks; 2) using IRB, we assess the performance of frontier LLMs and retrievers and gain interesting insights, such as the superior robustness of reasoning models in adversarial RAG scenarios and the critical importance of retriever quality for system improvement; and 3) we open-source our code and data to support future research.


\section{Related Work}

\noindent \textbf{Benchmarking factuality of LLMs.} IRB aligns with existing literature on assessing the ability of LLMs to answer factual questions. Early benchmarks in this domain include TriviaQA \cite{joshi2017triviaqa}, Natural Questions \cite{kwiatkowski2019natural}, HotpotQA \cite{yang2018hotpotqa}, 2WikiMultiHopQA \cite{ho2020constructing}, and FEVER \cite{thorne2018fever}. However, as model performance on these datasets has saturated and concerns regarding data contamination have grown, a new wave of benchmarks has emerged \cite{lin2022truthfulqa, niu2023ragtruth, li-etal-2023-halueval, vu2023freshllms, krishna2024fact, hsieh2024ruler, yang2024crag, wei2024long, wei2024measuring, pham2025sealqa, bang2025hallulens, choi2025citeguard}. While these recent efforts address distinct challenges, such as time-sensitive questions \cite{vu2023freshllms, yang2024crag, pham2025sealqa}, open-ended generation \cite{wei2024long}, and citation attribution \cite{choi2025citeguard}, they remain largely static or rely on manual updates. IRB distinguishes itself via an automatic generation pipeline that ensures meaningful evaluation by enabling on-demand updates.

\noindent \textbf{Automatic benchmark generation.} Prior work has explored automating the creation of RAG benchmarks to reduce reliance on manual annotation. Frameworks like RAGEval \cite{zhu2024rageval} and DataMorgana \cite{filice2025generating} rely on the instruction-following capabilities of LLMs to synthesize questions across various personas. Alternatively, FreshStack \cite{thakur2025freshstack} utilizes human-generated Q\&A pairs from StackOverflow, employing a search engine to fetch evidence which is subsequently validated by an LLM. In contrast, IRB grounds its generation in human-written citing sentences, where cited websites serve as evidence. This approach better reflects authentic human information-seeking patterns by leveraging documents that human writers explicitly selected as evidence. Furthermore, unlike methods that depend solely on LLM instruction-following, IRB achieves greater controllability in question-answer generation through a structured knowledge graph-based algorithm.

\section{Methodology}
\begin{figure*}[h]
\centering
\includegraphics[width=\textwidth]{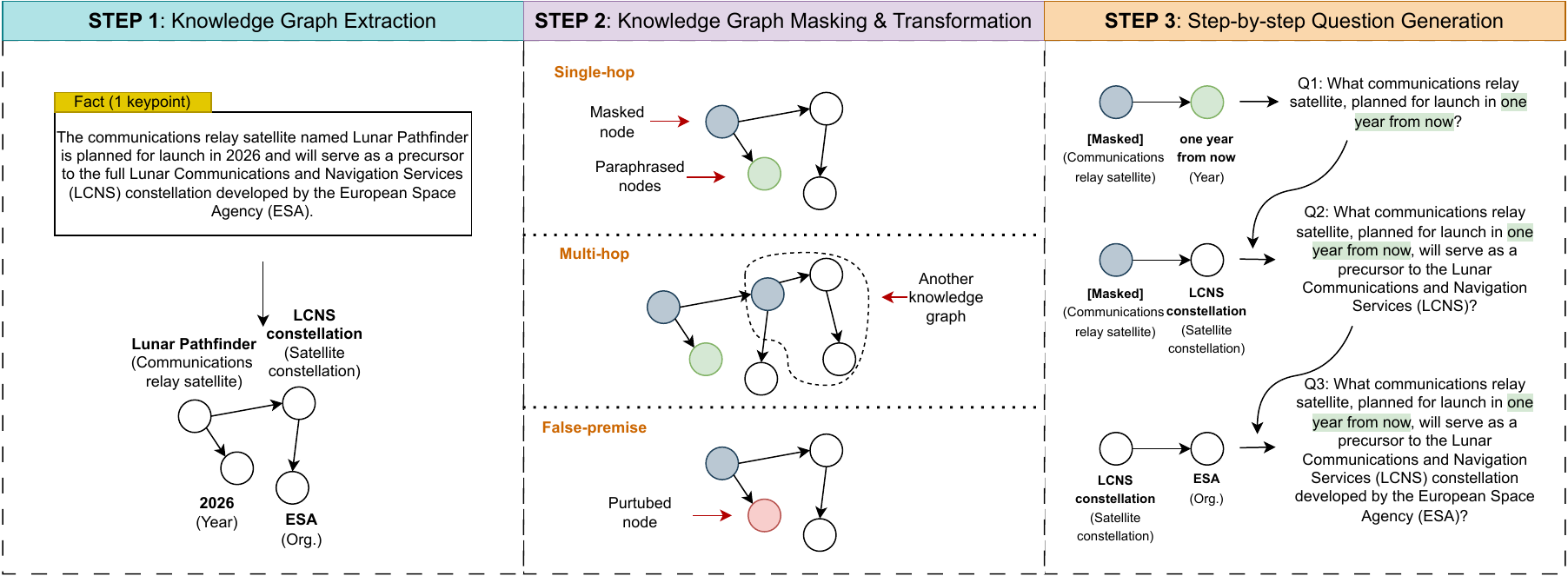}
\caption{The question generation process operates in three stages. First, a fact is structured into a knowledge graph. Subsequently, this graph is transformed into up to three distinct variants, namely single-hop, multi-hop, and false-premise. Finally, each variant is utilized to generate a corresponding natural language question. In this figure, we only show the question generation process for the single-hop variant. The reference date is 29 Sept. 2025.}
\label{fig:IRBQuestionGeneration}
\end{figure*}

IRB takes as input a collection of Wikipedia articles and outputs a benchmark dataset, containing a set of queries, their attributes, ground-truth answers, a reference corpus, and query relevance judgments (qrels). To ensure high-quality, factually-grounded benchmark dataset, we rely on two ``scaffolds'' to guardrail the benchmark generation process. First, we generate facts and supporting documents using human-written \textit{citing sentences} from Wikipedia as \textbf{factual scaffold}. These evidence-backed facts serve as the reliable base upon which question-answer (QA) pairs are generated. 

Then, we generate QA pairs from the evidence-backed facts using \textbf{algorithmic scaffold}. Specifically, we devised a knowledge graph-based question generation algorithm that takes a fact as input and generates high-quality question-answer pairs in a structured and controllable manner.

\subsection{Facts \& Supporting Documents Extraction}

Given a Wikipedia article, our pipeline begins by extracting \textit{citing sentences}. However, we observe that many extractions, such as cited section titles, lack informational value. To filter this noise, we apply a heuristic that retains only syntactically complete sentences, i.e., those containing a subject, predicate, and object.

Next, it is worth noting that a single citing sentence may contain multiple distinct claims, separated by the position of their inline citations. For example, in Figure \ref{fig:IRBFactAndSupportingDoc}, the citing sentence encompasses two claims, each supported by a different set of documents. Therefore, we split such sentences into distinct segments, each representing a more atomic fact. However, generating good facts requires more than just splitting the citing sentence into segments. The segment may be ambiguous (as exemplified by the two segments in Figure \ref{fig:IRBFactAndSupportingDoc}), making them unfit for generating well-formed questions and difficult to reliably verify against source text. To address this, we prompt an LLM to decontextualize the segments into a more self-contained version of themselves, which we call \textit{keypoints}. This step ensures the content is explicit enough for the subsequent groundedness check. For decontextualization, the LLM is provided with necessary information, including title and abstract of the source Wikipedia page, as well as the surrounding sentences. We provide the full prompt, which is inspired by \cite{gunjal2024molecular}, in Figure \ref{fig:keypoints_generation_prompt}.

Each keypoint is associated with a set of corresponding supporting documents, i.e., the set of URLs found within its original citation group. We retrieve the content, language, and publication date for each of these documents using \texttt{Trafilatura} \cite{barbaresi2021trafilatura}, \texttt{fast-langdetect} and \texttt{htmldate} \cite{barbaresi-2020-htmldate} respectively.

\noindent \textbf{Groundedness check}. It is natural to assume that all (keypoint, document) pairs are reliable because they originate from human-written citations. However, we find that a non-negligible portion of facts are not actually grounded in their retrieved documents. The primary cause is rarely human error but rather technical failures, including 1) availability issues (offline websites), 2) dynamic content leading to incomplete text retrieval, and 3) multimodal dependencies, where verification requires non-textual data (e.g., video or audio) beyond our current text-only pipeline. To address this, we verify all pairs by prompting an LLM to assess if the keypoint is supported, as shown in Figure \ref{fig:IRBFactAndSupportingDoc}. Only keypoints supported by at least one document are retained. Consequently, the final output is a \textit{Fact} composed of validated keypoints, each explicitly linked to its specific supporting evidence. The prompt for the groundedness check is provided in Figure \ref{fig:groundedness_check_prompt}.


\subsection{Generation of Question-Answer Pairs}

Unlike facts and supporting documents, which rely on the factual scaffold of human-written citing sentences, Question-Answer pairs are generated using an algorithmic scaffold. While a simplistic alternative would be to prompt an LLM to generate QA pairs directly from extracted facts, this approach offers \textit{little to no control over the generation process}. Specifically, direct prompting prevents us from systematically dictating the question's topic or calibrating its reasoning complexity (e.g., specifying the number of hops for multi-hop questions).

To address these limitations, we propose a graph-based generation pipeline, inspired by techniques in controllable question generation \cite{cheng2021guiding}. As illustrated in Figure \ref{fig:IRBQuestionGeneration}, this process proceeds in three distinct steps: \textit{knowledge graph construction}, \textit{knowledge graph masking \& transformation}, and \textit{step-by-step question generation}.

\subsubsection{Knowledge graph construction}

The first step in our algorithm is to convert each fact from unstructured text into a structured knowledge graph (KG). This graph is defined as a collection of triplets (head, relation, tail), along with the node type of each head and tail. For transparency, we provide the full prompt in Figure \ref{fig:knowledge_graph_extraction_prompt}.

To ensure the graph's content sufficiently covers the input answer, we validate it with a simple coverage check. This metric measures the proportion of words from the keypoints that are covered by the heads, relations, and tails of the extracted knowledge graph. Only graphs that exceed a certain coverage threshold are accepted.

\subsubsection{Knowledge graph masking \& transformation}
\label{sec:knowledge_graph_masking_transformation}

Next, we perform knowledge graph masking and transformation. This step dictates the answer and the complexity of the question to be generated. For each graph, we produce three masked versions corresponding to distinct question types namely \textit{single-hop}, \textit{multi-hop}, and \textit{false-premise}. 

\noindent \textbf{Single-hop.} In this setting, only one node is masked, which serves as the answer for the generated question. We select the node to mask by identifying the first \textit{maskable} head. A node is considered maskable if it satisfies specific criteria, such as being a named entity and appearing in all keypoints (we detail all criteria in \S \ref{sec:criteria_masked_node}).

\noindent \textbf{Multi-hop.} We construct masked graphs for multi-hop questions by merging two single-hop instances. Specifically, if the masked node of $KG_2$ appears as an unmasked node in $KG_1$, the two graphs are combined to form a composite graph $KG'$. This composite graph is used to generate a multi-hop question where the final answer aligns with the target of $KG_1$, but the reasoning process requires first resolving the sub-question implied by $KG_2$. Although we restrict our analysis to two-hop chains in this work, our method generalizes to longer reasoning paths.


\noindent \textbf{Node paraphrasing.} To increase question difficulty and prevent trivial solutions via exact keyword matching, we apply paraphrasing to unmasked nodes in both single-hop and multi-hop graphs. We employ rule-based transformations for specific node types, such as \textit{Person} (via name abbreviation, e.g., Cristiano Ronaldo $\to$ C. Ronaldo) and \textit{Date} (via relative time transformation, e.g., 30 November 2024 $\to$ 9 months ago\footnote{Calculated relative to a reference date of 29 September 2025.}). Table \ref{tab:transformations} details all paraphrasable node types and their corresponding transformation rules.

\noindent \textbf{False-premise generation.} Beyond paraphrasing, we also perturb nodes to inject false premises into the generated questions. To achieve this, we employ rule-based transformations on the same node types used for paraphrasing. For \textit{Person} nodes, we substitute the surname with a randomized alternative (e.g., Cristiano Ronaldo $\to$ C. Smith). Similarly, for \textit{Date} nodes, we apply a distorted relative time transformation (e.g., mapping 30 November 2024 $\to$ 1 year 4 months ago, intentionally deviating from the correct interval). We detail all transformable node types and their transformation rules in Table \ref{tab:pertubation_false_premise}.

\subsubsection{Step-by-step question generation}
We subsequently perform step-by-step question generation over the masked and transformed knowledge graph. We iteratively generate an intermediate question conditioned on the current triplet, as well as the history of previous triplets and generated questions. If the current triplet contains a masked node, the algorithm replaces the entity with a modified clause. Conversely, if no masking is present, the algorithm incorporates the triplet's information to refine the question's specificity. We provide the full prompt in Figure \ref{fig:question_generation_prompt}.

\noindent \textbf{Question answerability check \& refinement}. The generation of questions with multiple potential answers is an inevitable byproduct of the process. This arises not from limitations in the graph-guided generation algorithm, but from inherent ambiguities within the extracted facts. As these ambiguities can often be identified using general world knowledge, we prompt an LLM to assess whether a generated question admits a unique answer. Additionally, as stepwise generation may result in disjointed or verbose phrasing, we implement a final refinement module by prompting an LLM. We provide the full prompts in Figure \ref{fig:question_answerability_prompt} and \ref{fig:question_refinement_prompt}.

\subsection{Attributes}

Each sample generated by IRB is annotated with attributes derived from distinct sources. First, we extract evidence attributes, namely \textit{publication date} and \textit{language}, from supporting documents; these define the specific search and interpretation capabilities a system needs. Second, we record question characteristics, including \textit{number of hops} and \textit{false-premise status}, during the generation step to quantify the question's difficulty. Finally, we include the topic, mined from the source Wikipedia headers. These annotations enable the flexible construction of evaluation subsets. For example, to assess multi-hop reasoning in a cross-lingual setting, one can simply select samples whose questions are multi-hop and require non-English supporting evidence.

\subsection{Evaluation Protocols}


\noindent \textbf{Retriever evaluation metrics.} We adhere to standard text retrieval protocols, utilizing nDCG@$k$ \cite{jarvelin2002cumulated, jarvelin2017ir} to evaluate performance, where $k$ denotes the number of documents retrieved as context for answer generation. To establish relevance judgments (qrels), we define relevant documents as the evidence containing the facts from which a query was generated. This definition extends to false-premise queries, as the model requires the underlying factual evidence to correctly identify and refute the false premise.

\noindent \textbf{Generator evaluation metrics.} Given a query, the generator produces an answer either without external context (closed-book setting) or conditioned on the top $k$ retrieved documents (RAG setting). We evaluate the quality of these generations by comparing them to the ground truth. Following \cite{wei2024measuring, pham2025sealqa}, we employ an LLM-based evaluator to classify each prediction as ``correct'', ``incorrect'', or ``not attempted'', where we average the assessment of 2 LLMs (\texttt{GPT-4.1-mini} and \texttt{Qwen3-Next-80B} \cite{qwen3technicalreport}).

\section{Experiments}

\subsection{IRB1K}

\subsubsection{Implementation Details}

For our later experiments, we generated a RAG benchmark from a corpus of 2,000 recent Wikipedia articles (1,000 from 2024 and 1,000 from 2025) sourced from the September 29, 2025 Wikipedia dump. The pipeline yielded 1,838 questions in total. While the generation process is highly scalable (costing approximately \$18 and taking 16 hours using \texttt{GPT-4.1-mini}) we sampled 1,000 questions for our final evaluation set due to the high resource costs associated with evaluating frontier LLMs. For convenience, we name this dataset \texttt{IRB1K}.\footnote{We call API in sequence and not in parallel}. We provide the detailed statistics in Table \ref{tab:irb1k_statistics}.

\subsubsection{Quality Assessment}
\label{sec:irb1k_quality_assessment}

\begin{table}[]
\centering
\resizebox{\columnwidth}{!}{
\begin{tabular}{@{}llcc@{}}
\toprule
\textbf{Outcome Category}     & \textbf{Specific Criterion} & \multicolumn{1}{l}{\textbf{\# Samples}} & \multicolumn{1}{l}{\textbf{Percentage (\%)}} \\ \midrule
Success                       & All criteria satisfied      & 189                                     & 94.5                                         \\
\multirow{3}{*}{Failure Mode} & 1. Question Malformed       & 10                                      & 5                                            \\
                              & 2. Fact Unnecessary         & 0                                       & 0                                            \\
                              & 3. Answer Invalid           & 1                                       & 0.5                                          \\
Total                         &                             & 200                                     & 100                                          \\ \bottomrule
\end{tabular}
}

\caption{Quality assessment results.}
\label{tab:irb1k_quality_assessment}
\end{table}

To validate the quality of the dataset generated by our proposed framework, we conduct a manual evaluation on 200 randomly sampled QA pairs. The evaluation was performed by volunteer undergraduate and graduate students in Computer Science and Electrical Engineering in the United States. To ensure accuracy, we ask two annotators to first independently assess each sample; then, in case of disagreement, they discussed the sample to reach a final decision. Each sample was evaluated against three criteria in the following order\footnote{The full rubric is provided in \S \ref{sec:human_eval_rubrics_interface}}.

\begin{table*}[h]
\centering
\resizebox{0.85\textwidth}{!}{

\begin{tabular}{@{}lccccccccccc@{}}
\toprule
\multicolumn{1}{l|}{\multirow{3}{*}{}}      & \multicolumn{10}{c|}{\textbf{Valid-premise}}                                                                                                                                                                                                      & \multirow{3}{*}{\textbf{False-premise}} \\ \cmidrule(lr){2-11}
\multicolumn{1}{l|}{}                       & \multicolumn{2}{c|}{\textbf{Language}}             & \multicolumn{2}{c|}{\textbf{Freshness}}            & \multicolumn{4}{c|}{\textbf{Topic}}                                                & \multicolumn{2}{c|}{\textbf{\# Hops}}              &                                         \\ \cmidrule(lr){2-11}
\multicolumn{1}{l|}{}                       & English       & \multicolumn{1}{c|}{Cross}         & 2024          & \multicolumn{1}{c|}{2025}          & Culture       & Geo           & H \& S        & \multicolumn{1}{c|}{STEM}          & Single        & \multicolumn{1}{c|}{Multi}         &                                         \\ \midrule
\multicolumn{12}{c}{\textbf{First stage retrieval}}                                                                                                                                                                                                                                                                                       \\ \midrule
\multicolumn{1}{l|}{BM25}                   & \textbf{75.6} & \multicolumn{1}{c|}{13.4}          & 58.4          & \multicolumn{1}{c|}{54.1}          & 52.2          & 50.3          & {\ul 75.5}    & \multicolumn{1}{c|}{\textbf{80.3}} & 56            & \multicolumn{1}{c|}{\textbf{56.7}} & 56.4                                    \\
\multicolumn{1}{l|}{text-embedding-3-small} & 69.4          & \multicolumn{1}{c|}{{\ul 50.2}}    & {\ul 66.5}    & \multicolumn{1}{c|}{{\ul 60.7}}    & {\ul 61.3}    & {\ul 61.6}    & 73.6          & \multicolumn{1}{c|}{{\ul 78.9}}    & {\ul 64.5}    & \multicolumn{1}{c|}{49.9}          & {\ul 59.2}                              \\
\multicolumn{1}{l|}{E5-base-v2}             & 67.9          & \multicolumn{1}{c|}{34}            & 63            & \multicolumn{1}{c|}{52.4}          & 53.4          & 54.7          & 72.6          & \multicolumn{1}{c|}{75.5}          & 57.3          & \multicolumn{1}{c|}{56.2}          & 56.1                                    \\
\multicolumn{1}{l|}{BGE-m3}                 & {\ul 69.5}    & \multicolumn{1}{c|}{\textbf{53.6}} & \textbf{68.9} & \multicolumn{1}{c|}{\textbf{60.8}} & \textbf{61.7} & \textbf{63}   & \textbf{75.9} & \multicolumn{1}{c|}{78.4}          & \textbf{65.2} & \multicolumn{1}{c|}{{\ul 56.6}}    & \textbf{66.5}                           \\ \midrule
\multicolumn{12}{c}{\textbf{Reranking with Cohere Rerank 3.5}}                                                                                                                                                                                                                                                                            \\ \midrule
\multicolumn{1}{l|}{BM25}                   & \textbf{83.6} & \multicolumn{1}{c|}{32.8}          & 72.1          & \multicolumn{1}{c|}{63.8}          & 65.2          & 62.7          & 83.3          & \multicolumn{1}{c|}{88.9}          & 67.7          & \multicolumn{1}{c|}{66.4}          & 68.4                                    \\
\multicolumn{1}{l|}{text-embedding-3-small} & {\ul 82.8}    & \multicolumn{1}{c|}{{\ul 67.9}}    & {\ul 81.7}    & \multicolumn{1}{c|}{{\ul 75.1}}    & {\ul 75.8}    & {\ul 75.8}    & {\ul 85.6}    & \multicolumn{1}{c|}{\textbf{91.6}} & {\ul 78.9}    & \multicolumn{1}{c|}{68.6}          & {\ul 79.3}                              \\
\multicolumn{1}{l|}{E5-base-v2}             & 82.2          & \multicolumn{1}{c|}{60}            & 78.4          & \multicolumn{1}{c|}{72.5}          & 72.6          & 72.1          & 84.8          & \multicolumn{1}{c|}{\textbf{91.6}} & 75.7          & \multicolumn{1}{c|}{{\ul 69.2}}    & 76.4                                    \\
\multicolumn{1}{l|}{BGE-m3}                 & 82.5          & \multicolumn{1}{c|}{\textbf{73.1}} & \textbf{82.8} & \multicolumn{1}{c|}{\textbf{76.8}} & \textbf{77.1} & \textbf{77.6} & \textbf{85.8} & \multicolumn{1}{c|}{91.1}          & \textbf{80.3} & \multicolumn{1}{c|}{\textbf{70.4}} & \textbf{80.2}                           \\ \bottomrule
\end{tabular}

}

\caption{Performance (nDCG@5) of retrievers on IRB1K. We \textbf{bold} and \underline{underline} the best and second-best results in each setting. H \& S is short for History \& Society}
\label{tab:retrieval_performance_irb1k}
\end{table*}

\noindent \textbf{Question well-formedness}: The question must be specific and understandable. Ill-formed questions can confuse the RAG system into finding the wrong answer, leading to unreliable evaluation signals.

\noindent \textbf{Fact necessity}: Answering the question must require integrating information from all parts of the compositional fact. This assesses if retrieval of all supporting evidence is necessary.

\noindent \textbf{Answer validity}: The ground-truth answer must be correct and unique. Conversely, when an answer is wrong or not unique (is one of the possible answers), evaluation of RAG systems would be unreliable.

Table \ref{tab:irb1k_quality_assessment} presents the evaluation results, which strongly validate the IRB framework. Specifically, approximately 95\% of the evaluated samples satisfy all three criteria. Analysis of the remaining 5\% reveals that the majority of errors stem from malformed questions attributed to the knowledge graph construction process (e.g., misidentified node types or information loss during conversion). Only a single instance exhibited the issue of multiple answers.


\subsection{Experiment Setup}

In our experiments, we assess a straightforward RAG pipeline with a simple prompt similar to \cite{yang2024crag}, where we also encourage brief answers and saying ``I don't know'' or ``False premise question'' (see Figure \ref{fig:question_answering_prompt_with_retrieval} and \ref{fig:question_answering_prompt_without_retrieval}). Unless specified otherwise, for each query, we retrieve the top $k=5$ documents and formatted them similarly to \texttt{FRESH-PROMPT} \cite{vu2023freshllms}, where aside from the main content, we include the source webpage and the publication date. Crucially, we set the ``current date'' in the prompt to 29 September 2025, corresponding to our Wikipedia dump date. Reported results represent a single run.

\noindent \textbf{Retriever}. For our main analysis, we employ OpenAI's \texttt{text-embedding-3-small} as our primary retriever. However, we do evaluate other retrievers, namely \texttt{BM25} \cite{robertson2009probabilistic}, \texttt{BGE-M3} \cite{chen2024bge} and \texttt{E5-base} \cite{wang2022text}. The documents in the corpus are split into chunks of 512 tokens. During inference, we retrieve the top 5 documents, where their rankings are determined according to their highest ranked chunk, a setting similar to \cite{kaszkiel1997passage}.

\noindent \textbf{Generator}. We evaluate a diverse set of both proprietary and open-source LLMs. For reasoning-focused models, these include proprietary models namely \texttt{GPT-5-mini} and \texttt{GPT-5} \cite{openai2025gpt5}, as well as leading open-source models like \texttt{gpt-oss-120b} \cite{openai2025gpt} and \texttt{DeepSeek-R1} \cite{guo2025deepseek}. We utilize the medium reasoning mode for GPT models and allocate a reasoning budget of 2048 tokens for the others. For comparison, we also evaluate non-reasoning models, including proprietary models that are \texttt{GPT-4.1-mini}, \texttt{GPT-4.1} \cite{openai2025gpt41}, and open-source models namely \texttt{Llama-3.3-70B} \cite{meta2025llama33} and \texttt{Llama-4-Scout} \cite{meta2025llama4}.

\subsection{Evaluation Results}
\begin{table*}[]
\centering

\resizebox{0.85\textwidth}{!}{

\begin{tabular}{@{}lccccccccccc@{}}
\toprule
\multicolumn{1}{l|}{\multirow{3}{*}{}} & \multicolumn{10}{c|}{\textbf{Valid-premise}}                                                                                                                                                                                                      & \multirow{3}{*}{\textbf{False-premise}} \\ \cmidrule(lr){2-11}
\multicolumn{1}{l|}{}                  & \multicolumn{2}{c|}{\textbf{Language}}             & \multicolumn{2}{c|}{\textbf{Freshness}}            & \multicolumn{4}{c|}{\textbf{Topic}}                                                & \multicolumn{2}{c|}{\textbf{\# Hops}}              &                                         \\ \cmidrule(lr){2-11}
\multicolumn{1}{l|}{}                  & English       & \multicolumn{1}{c|}{Cross}         & 2024          & \multicolumn{1}{c|}{2025}          & Culture       & Geo           & H \& S        & \multicolumn{1}{c|}{STEM}          & Single        & \multicolumn{1}{c|}{Multi}         &                                         \\ \midrule
\multicolumn{12}{c}{\textbf{Closed-book performance}}                                                                                                                                                                                                                                                                                \\ \midrule
\multicolumn{1}{l|}{GPT-4.1 mini}      & 11.4          & \multicolumn{1}{c|}{9.7}           & 11.5          & \multicolumn{1}{c|}{10.4}          & 5.9           & 10.4          & 20.3          & \multicolumn{1}{c|}{22.7}          & 11.1          & \multicolumn{1}{c|}{8.1}           & 10                                      \\
\multicolumn{1}{l|}{GPT-4.1}           & {\ul 30.6}    & \multicolumn{1}{c|}{{\ul 21.4}}    & {\ul 32}      & \multicolumn{1}{c|}{{\ul 24.1}}    & {\ul 18.4}    & {\ul 26.1}    & {\ul 40.4}    & \multicolumn{1}{c|}{{\ul 43.2}}    & {\ul 28.3}    & \multicolumn{1}{c|}{{\ul 20.2}}    & 1                                       \\
\multicolumn{1}{l|}{GPT-5 mini medium} & 14.8          & \multicolumn{1}{c|}{9.1}           & 16.1          & \multicolumn{1}{c|}{10.4}          & 5.8           & 10.8          & 26.2          & \multicolumn{1}{c|}{28.2}          & 12.9          & \multicolumn{1}{c|}{14.5}          & 8                                       \\
\multicolumn{1}{l|}{GPT-5 medium}      & \textbf{41.9} & \multicolumn{1}{c|}{\textbf{31.5}} & \textbf{48.4} & \multicolumn{1}{c|}{\textbf{30.4}} & \textbf{29.8} & \textbf{36.8} & \textbf{54.9} & \multicolumn{1}{c|}{\textbf{58.2}} & \textbf{39.4} & \multicolumn{1}{c|}{\textbf{29}}   & 24                                      \\
\multicolumn{1}{l|}{Llama-4-Scout}     & 8.5           & \multicolumn{1}{c|}{4.8}           & 10            & \multicolumn{1}{c|}{5.1}           & 3.2           & 5.1           & 14            & \multicolumn{1}{c|}{17.3}          & 7.7           & \multicolumn{1}{c|}{3.2}           & {\ul 31}                                \\
\multicolumn{1}{l|}{Llama-3.3-70B}     & 14.8          & \multicolumn{1}{c|}{9.5}           & 16.7          & \multicolumn{1}{c|}{10.1}          & 4.9           & 9.5           & 29            & \multicolumn{1}{c|}{34.5}          & 13.1          & \multicolumn{1}{c|}{12.9}          & 12.2                                    \\
\multicolumn{1}{l|}{gpt-oss-120b}      & 16.3          & \multicolumn{1}{c|}{11.9}          & 16.1          & \multicolumn{1}{c|}{13.9}          & 5.7           & 14.3          & 28            & \multicolumn{1}{c|}{28.2}          & 15.4          & \multicolumn{1}{c|}{8.9}           & 11.5                                    \\
\multicolumn{1}{l|}{DeepSeek-R1}       & 11.3          & \multicolumn{1}{c|}{8.5}           & 13.4          & \multicolumn{1}{c|}{7.9}           & 7.5           & 9.1           & 14.7          & \multicolumn{1}{c|}{17.7}          & 10.9          & \multicolumn{1}{c|}{4.8}           & \textbf{39.5}                           \\ \midrule
\multicolumn{12}{c}{\textbf{RAG performance}}                                                                                                                                                                                                                                                                                        \\ \midrule
\multicolumn{1}{l|}{GPT-4.1 mini}      & 82.8          & \multicolumn{1}{c|}{72.2}          & 76.8          & \multicolumn{1}{c|}{81.7}          & 77.9          & 79.1          & 85            & \multicolumn{1}{c|}{81.8}          & 80.6          & \multicolumn{1}{c|}{66.1}          & 1.5                                     \\
\multicolumn{1}{l|}{GPT-4.1}           & \textbf{84.9} & \multicolumn{1}{c|}{{\ul 73.4}}    & {\ul 80.1}    & \multicolumn{1}{c|}{{\ul 82.4}}    & {\ul 79.1}    & {\ul 80.6}    & \textbf{86.4} & \multicolumn{1}{c|}{84.5}          & {\ul 82.5}    & \multicolumn{1}{c|}{66.9}          & 2.8                                     \\
\multicolumn{1}{l|}{GPT-5 mini medium} & 83.1          & \multicolumn{1}{c|}{72.8}          & 77.6          & \multicolumn{1}{c|}{81.8}          & 77.4          & 80.1          & 84.8          & \multicolumn{1}{c|}{82.3}          & 81            & \multicolumn{1}{c|}{66.1}          & {\ul 35}                                \\
\multicolumn{1}{l|}{GPT-5 medium}      & \textbf{84.9} & \multicolumn{1}{c|}{\textbf{77.6}} & \textbf{81.4} & \multicolumn{1}{c|}{\textbf{83.5}} & \textbf{80.9} & \textbf{82.6} & 85            & \multicolumn{1}{c|}{{\ul 85.5}}    & \textbf{83.5} & \multicolumn{1}{c|}{{\ul 71}}      & \textbf{50.5}                           \\
\multicolumn{1}{l|}{Llama-4-Scout}     & 76            & \multicolumn{1}{c|}{62.9}          & 71.9          & \multicolumn{1}{c|}{71.9}          & 70.2          & 70.3          & 75.9          & \multicolumn{1}{c|}{80}            & 73.5          & \multicolumn{1}{c|}{52.4}          & 20.5                                    \\
\multicolumn{1}{l|}{Llama-3.3-70B}     & 78.6          & \multicolumn{1}{c|}{62.9}          & 73.6          & \multicolumn{1}{c|}{73.6}          & 72.7          & 71.3          & 80.4          & \multicolumn{1}{c|}{80.9}          & 75.2          & \multicolumn{1}{c|}{54.8}          & 5                                       \\
\multicolumn{1}{l|}{gpt-oss-120b}      & 84.4          & \multicolumn{1}{c|}{69.8}          & 79.9          & \multicolumn{1}{c|}{79.7}          & 78.3          & 78            & {\ul 86.2}    & \multicolumn{1}{c|}{\textbf{86.4}} & 80.4          & \multicolumn{1}{c|}{\textbf{73.4}} & 14.5                                    \\
\multicolumn{1}{l|}{DeepSeek-R1}       & 79.9          & \multicolumn{1}{c|}{71.4}          & 76.2          & \multicolumn{1}{c|}{78.1}          & 75.1          & 77.6          & 80.6          & \multicolumn{1}{c|}{80.5}          & 78.5          & \multicolumn{1}{c|}{62.1}          & 22.8                                    \\ \bottomrule
\end{tabular}

}

\caption{Evaluation of LLM \textbf{correctness} on IRB1K. Results indicate that while RAG bridges the performance gap present in closed-book settings, models continue to struggle with specific complexities. Notably, performance degrades in cross-lingual, fresh information, multi-hop, and false-premise scenarios. We \textbf{bold} and \underline{underline} the best and second-best results in each setting.}
\label{tab:correctness_irb1k}

\vspace{-0.5cm}
\end{table*}
\subsubsection{The necessity of retrieval for IRB1K}
Firstly, we would like to ensure that questions generated by IRB are not trivial and can be answered without access to search results. As illustrated in Table \ref{tab:correctness_irb1k}, IRB1K presents a substantial challenge to frontiers LLMs when operating without access to external information. While models with more recent knowledge cutoffs (e.g. GPT-5 and GPT-5 mini) exhibit marginal improvements, the overall low performance confirms that our framework successfully generates non-trivial queries. 


\subsubsection{Effectiveness of retrievers}

Table \ref{tab:retrieval_performance_irb1k} summarizes the effectiveness of retrievers on IRB1K. Unsurprisingly, we observe that effectiveness of retrievers varies across queries of different attributes. Specifically, \textit{retrievers struggle with questions 1) in the multilingual setting, where an English question is used to search for supporting documents in other languages; 2) involving fresh information; 3) with multiple reasoning hops.} Interestingly, \textit{topic-based analysis indicates that retrievers remain most effective within the STEM domain, while struggling to maintain similar ranking performance in other topical categories.} Furthermore, BM25 remains a robust baseline, as it outperforms all neural retrievers on monolingual English tasks, though it naturally underperforms in the cross-lingual setting.


\subsubsection{Effectiveness of RAG}

As shown in Table \ref{tab:correctness_irb1k}, the correctness of all evaluated LLMs on IRB1K improves substantially when provided with retrieved context.

\noindent \textbf{Retrieval acts as an ``equalizer''}. In the absence of retrieval, performance disparities between LLMs are pronounced. Specifically, the gap between the highest (\texttt{GPT-5}) and lowest-performing (\texttt{Llama-4-Scout}) models is nearly four-fold. Furthermore, individual model performance fluctuates across different question types. However, when retrieval is integrated, the performance gap between models shrinks significantly, and all models exhibit greater consistency across tasks.

\begin{table}[]
\centering
\resizebox{0.9\columnwidth}{!}{

\begin{tabular}{@{}l|ccc|ccc@{}}
\toprule
                   & \multicolumn{3}{c|}{\textbf{Correct retrieval}}                                                             & \multicolumn{3}{c}{\textbf{Incorrect retrieval}}                                                             \\ \cmidrule(l){2-7} 
\multirow{-2}{*}{} & \cellcolor[HTML]{D9EAD3}C             & \cellcolor[HTML]{F4CCCC}I             & \cellcolor[HTML]{FFF2CC}N   & \cellcolor[HTML]{D9EAD3}C             & \cellcolor[HTML]{F4CCCC}I             & \cellcolor[HTML]{FFF2CC}N    \\ \midrule
GPT-4.1 mini       & \cellcolor[HTML]{D9EAD3}71.8          & \cellcolor[HTML]{F4CCCC}26.3          & \cellcolor[HTML]{FFF2CC}1.9 & \cellcolor[HTML]{D9EAD3}38.4          & \cellcolor[HTML]{F4CCCC}35.3          & \cellcolor[HTML]{FFF2CC}26.3 \\
GPT-4.1            & \cellcolor[HTML]{D9EAD3}74.2          & \cellcolor[HTML]{F4CCCC}22.4          & \cellcolor[HTML]{FFF2CC}3.4 & \cellcolor[HTML]{D9EAD3}38.4          & \cellcolor[HTML]{F4CCCC}31            & \cellcolor[HTML]{FFF2CC}30.6 \\
GPT-5 mini medium  & \cellcolor[HTML]{D9EAD3}{\ul 79.4}    & \cellcolor[HTML]{F4CCCC}{\ul 16.6}    & \cellcolor[HTML]{FFF2CC}4   & \cellcolor[HTML]{D9EAD3}{\ul 45.3}    & \cellcolor[HTML]{F4CCCC}\textbf{15.4} & \cellcolor[HTML]{FFF2CC}39.3 \\
GPT-5 medium       & \cellcolor[HTML]{D9EAD3}\textbf{82.7} & \cellcolor[HTML]{F4CCCC}\textbf{15.5} & \cellcolor[HTML]{FFF2CC}1.8 & \cellcolor[HTML]{D9EAD3}\textbf{55.4} & \cellcolor[HTML]{F4CCCC}{\ul 18.1}    & \cellcolor[HTML]{FFF2CC}26.5 \\
Llama-4-Scout      & \cellcolor[HTML]{D9EAD3}70.2          & \cellcolor[HTML]{F4CCCC}25.7          & \cellcolor[HTML]{FFF2CC}4.1 & \cellcolor[HTML]{D9EAD3}33.3          & \cellcolor[HTML]{F4CCCC}36.6          & \cellcolor[HTML]{FFF2CC}30.1 \\
Llama-3.3-70B      & \cellcolor[HTML]{D9EAD3}69.7          & \cellcolor[HTML]{F4CCCC}20.3          & \cellcolor[HTML]{FFF2CC}10  & \cellcolor[HTML]{D9EAD3}29.7          & \cellcolor[HTML]{F4CCCC}18.3          & \cellcolor[HTML]{FFF2CC}52   \\
gpt-oss-120b       & \cellcolor[HTML]{D9EAD3}74.4          & \cellcolor[HTML]{F4CCCC}22.1          & \cellcolor[HTML]{FFF2CC}3.5 & \cellcolor[HTML]{D9EAD3}41.3          & \cellcolor[HTML]{F4CCCC}25.2          & \cellcolor[HTML]{FFF2CC}33.5 \\
DeepSeek-R1        & \cellcolor[HTML]{D9EAD3}76.2          & \cellcolor[HTML]{F4CCCC}20            & \cellcolor[HTML]{FFF2CC}3.8 & \cellcolor[HTML]{D9EAD3}35.7          & \cellcolor[HTML]{F4CCCC}25.7          & \cellcolor[HTML]{FFF2CC}38.6 \\ \bottomrule
\end{tabular}

}

\caption{Effectiveness of RAG systems when retriever is correct and incorrect. C, I, N are short for ``correct'', ``incorrect'' and ``not attempted'' respectively}
\label{tab:correct_incorrect}
\end{table}

\noindent \textbf{RAG correctness is bottlenecked by the retriever.} Table \ref{tab:correct_incorrect} compares system effectiveness under correct versus incorrect retrieval conditions. We observe that when retrieval is accurate, i.e. all gold documents are retrieved within the top-5 contexts, all LLMs achieve high performance. Conversely, incorrect retrieval causes a sharp decline in overall system correctness. Crucially, however, this drop in performance is not due to frontier LLMs producing more hallucinations. Rather, these models often successfully detect irrelevant contexts and refuse to answer, as evidenced by the increase in ``not attempted'' rates. Furthermore, as discussed in \S \ref{sec:knowledge_interplay}, when faced with incorrect retrieval results, models possessing the correct internal information effectively prioritize their parametric knowledge over the provided context. These observations indicate that frontier LLMs are already highly reliable readers. Thus, \textit{future improvements in RAG systems should prioritize the retrieval component over the generator.}

\noindent \textbf{False-premise questions pose a persistent challenge}. We observe that all models struggle to identify false-premise questions, particularly in the absence of retrieval. While \texttt{Llama-4-Scout} and \texttt{DeepSeek-R1} appear to be exceptions with higher detection rates, this is actually a result of model bias, as these models exhibit a tendency to label \textit{any} query as a ``False-premise question'', resulting in a disproportionately high false positive rate.

\noindent \textbf{Reasoning models are more reliable}. Reasoning models not only exhibit higher correctness but also generate fewer incorrect answers, as observed in Table \ref{tab:correctness_irb1k} and \ref{tab:correct_incorrect}. Furthermore, reasoning models are significantly more reliable in handling incorrect retrieval contexts. This is evidenced by their lower error rates when retrieval is wrong, compared to their non-reasoning counterparts. Furthermore, while all models struggle to detect false-premise questions, reasoning LLMs are generally more reliable.


\begin{figure}[h]
\centering
\includegraphics[width=0.9\columnwidth]{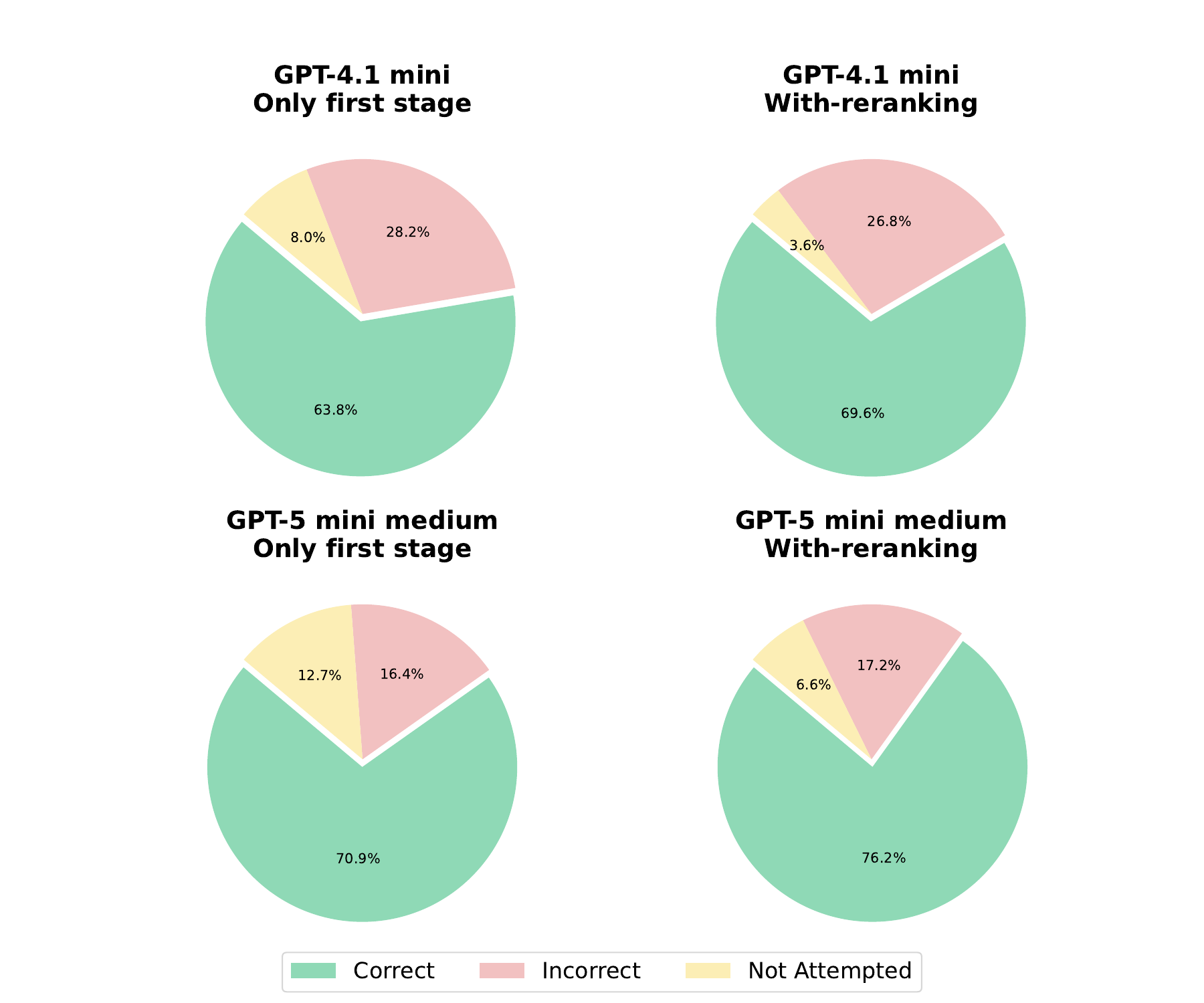}
\caption{Effect of reranking.}
\label{fig:EffectOfRerank}
\end{figure}

\noindent \textbf{Reranking enhances performance.} We investigate how reranking retrieval results from \texttt{text-embedding-3-small} impacts the effectiveness of RAG systems. As expected, we find that using reranked retrieval contexts consistently leads to gains in LLM correctness. This supports our earlier argument that future development should focus heavily on the retriever side.

\noindent \textbf{Additional experiment results.} Due to space limitations, we provide detailed supplementary analyses in the appendices. \S\ref{sec:reasoning_efforts_analysis} examines reasoning token allocation, revealing that reasoning LLMs expend more tokens on multi-hop questions and in closed-book settings. \S\ref{sec:knowledge_interplay} investigates the interplay between models' internal knowledge and retrieved evidence, demonstrating that models remain robust even when conflicts exist between the two sources.




\section{Conclusion}

We introduce IRB, a framework for automatically generating benchmarks for evaluating RAG systems. IRB automates the benchmark generation process through structured and scaffolded generation to produce high-quality datasets that reliably test RAG systems. Overall, our evaluations show that: 1) IRB generates challenging benchmarks that challenge frontier LLMs without access to external knowledge; 2) frontier LLMs, especially reasoning models, are reliable; and 3) retrievers should receive more attention, as improving retrieval is essential for making RAG systems more robust.

\section*{Limitations}

\noindent \textbf{Restriction to Short Answers.} IRB is designed to generate questions that elicit short, specific answers, following the approach of \cite{wei2024measuring, pham2025sealqa}. Consequently, benchmarks produced by our framework are not suitable for evaluating long-form text generation capabilities.

\noindent \textbf{Restriction to Wikipedia.} As noted, IRB relies on human-written sentences from Wikipedia to generate questions, answers, and evidence. While other domains (i.e. scientific literature and legal case law) also contain rich citation structures, the semantic function of their citations differs fundamentally from that of Wikipedia. Consequently, adapting IRB to these domains would require a substantial redesign of the pipeline. We therefore leave the extension of our framework to these alternative data sources for future work.



\section*{Ethical Considerations}

While we apply strict filtering to remove low-quality or harmful content, the IRB benchmark is derived from and represents the current state of Wikipedia. As such, we cannot guarantee the complete exclusion of sensitive information or copyrighted material within the questions and references. Researchers should use the dataset with the understanding that it mirrors the characteristics and potential liabilities of its source corpus.



\bibliography{custom}

\appendix
\label{sec:appendix}

\begin{table}[]
\centering
\resizebox{0.85\columnwidth}{!}{

\begin{tabular}{@{}ccc|c@{}}
\toprule
\multicolumn{3}{c|}{\textbf{Attribute of sample}}                                                                & \textbf{Percentage} \\ \midrule
\multicolumn{1}{c|}{\multirow{10}{*}{Valid-premise}} & \multicolumn{1}{c|}{\multirow{2}{*}{Language}}  & English & 54.8                \\
\multicolumn{1}{c|}{}                                & \multicolumn{1}{c|}{}                           & Cross   & 25.2                \\ \cmidrule(l){2-4} 
\multicolumn{1}{c|}{}                                & \multicolumn{1}{c|}{\multirow{2}{*}{Freshness}} & 2024    & 36.6                \\
\multicolumn{1}{c|}{}                                & \multicolumn{1}{c|}{}                           & 2025    & 43.4                \\ \cmidrule(l){2-4} 
\multicolumn{1}{c|}{}                                & \multicolumn{1}{c|}{\multirow{4}{*}{Topics}}    & Culture & 51.4                \\
\multicolumn{1}{c|}{}                                & \multicolumn{1}{c|}{}                           & Geo     & 49.2                \\
\multicolumn{1}{c|}{}                                & \multicolumn{1}{c|}{}                           & H \& S  & 21.4                \\
\multicolumn{1}{c|}{}                                & \multicolumn{1}{c|}{}                           & STEM    & 11                  \\ \cmidrule(l){2-4} 
\multicolumn{1}{c|}{}                                & \multicolumn{1}{c|}{\multirow{2}{*}{\# Hops}}   & Single  & 73.8                \\
\multicolumn{1}{c|}{}                                & \multicolumn{1}{c|}{}                           & Multi   & 6.2                 \\ \midrule
\multicolumn{3}{c|}{False-premise}                                                                               & 20                  \\ \bottomrule
\end{tabular}

}
\caption{IRB1K statistics. H \& S is short for History \& Society.}
\label{tab:irb1k_statistics}
\end{table}

\section{Additional Details of IRB}

\subsection{Criteria for selecting masked node}
\label{sec:criteria_masked_node}
As discussed in \S \ref{sec:knowledge_graph_masking_transformation}, when masking a knowledge graph, a candidate node must satisfy a specific set of criteria, which we detailed below. 

\begin{itemize}[leftmargin=*, noitemsep, topsep=0pt]
    \item \textbf{Named Entity}: The node must be identified as a named entity. We utilize \texttt{spaCy} \cite{honnibal2020spacy} to verify this property.
    \item \textbf{Coverage}: The node must appear in all keypoints associated with the fact.
    \item \textbf{Atomicity}: The node must not reference multiple entities. We validate this by ensuring the node text does not contain conjunctions such as ``and'', ``\&'', or ``et al''.
    \item \textbf{Non-overlapping}: The node text must not be contained within other nodes. We employ fuzzy matching to enforce this constraint.
    \item \textbf{Non-exclusive}: The node must serve as the unique head and tail for any associated relation.
\end{itemize}

\subsection{Entity transformation}

\subsection{Full prompts}
\noindent \textbf{Question generation}. We provide the full prompts used in the question generation process. Specifically, Figure \ref{fig:knowledge_graph_extraction_prompt} details the knowledge graph extraction prompt; Figure \ref{fig:question_generation_prompt} details the step-by-step question generation prompt; Figure \ref{fig:question_answerability_prompt} details the question answerability check; Figure \ref{fig:question_refinement_prompt} details the question refinement prompt.

\noindent \textbf{Question answering.} We provide the full prompts for generating answers in Figures \ref{fig:question_answering_prompt_with_retrieval} (with retrieval context) and \ref{fig:question_answering_prompt_without_retrieval} (without retrieval context).

\noindent \textbf{LLM-based evaluation.} We adopt the LLM-based evaluation prompt from \cite{pham2025sealqa}. This prompt instructs an LLM to classify an answer as ``correct'', ``incorrect'', or ``not attempted''.

\section{Additional Experimental Results}

\subsection{Varying number of retrieval contexts}

\begin{figure}[h]
\centering
\includegraphics[width=\columnwidth]{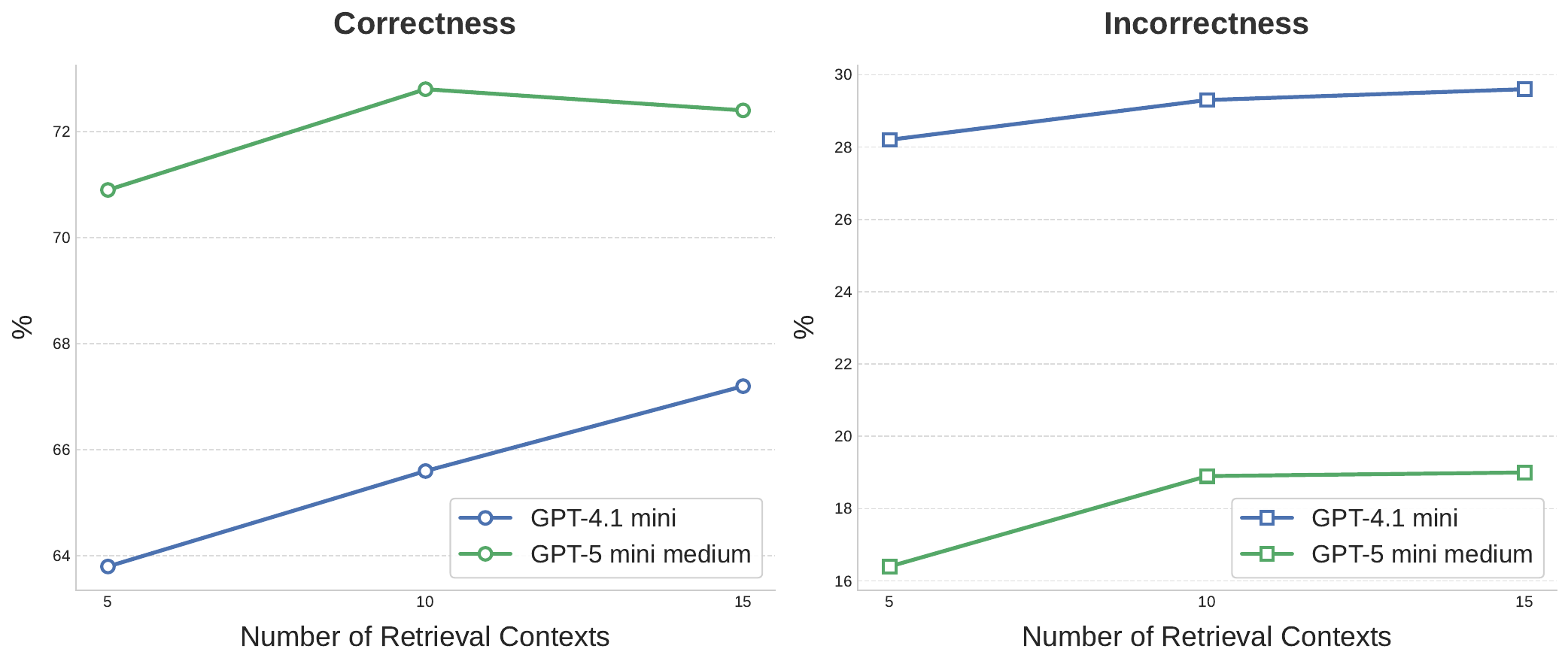}
\caption{RAG performance (correctness and incorrectness) when the number of retrieval contexts are increased}
\label{fig:VaryingRetrievalContexts}
\end{figure}

\label{sec:varying_num_retrieval_contexts}

Expanding the retrieval set presents a fundamental trade-off. On one hand, it maximizes the likelihood of capturing the gold document within the context window. On the other, it introduces noise and increases computational costs. We investigate how effectively LLMs utilize these expanded retrieval contexts. Given the computational costs of extensive evaluation, we restrict this analysis to two representative models, namely \texttt{GPT-4.1-mini} and \texttt{GPT-5-mini}, representing standard chat and reasoning models, respectively. Results are shown in Figure \ref{fig:VaryingRetrievalContexts}.

We observe that increasing the number of retrieved documents leads to a simultaneous rise in both correct and incorrect answers. Ideally, additional context should improve correctness while suppressing incorrectness. However, both models fail to demonstrate this behavior. The concurrent increase in positive and negative outcomes indicates that models become less likely to refuse to answer as context grows. This suggests an undesirable over-confidence, particularly when the expanded retrieval set fails to contain the correct document.

\subsection{Reasoning efforts analysis}

\label{sec:reasoning_efforts_analysis}

\begin{figure*}[h]
    \centering
    \begin{subfigure}[b]{\textwidth}
        \centering
        \includegraphics[width=\linewidth]{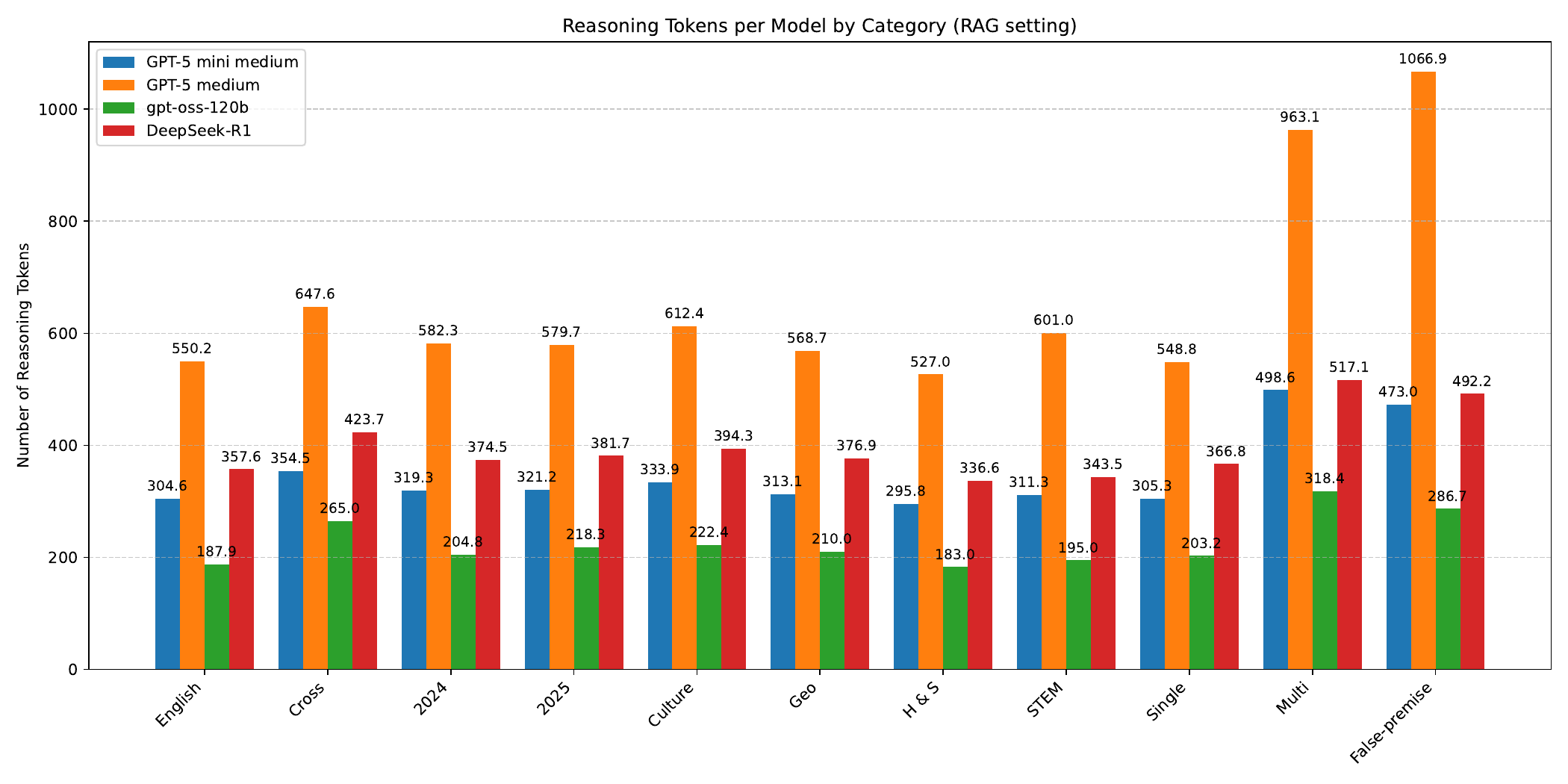}
        \label{fig:ReasoningEffortsVisRAG}
    \end{subfigure}
    
    \par\bigskip 
    
    \begin{subfigure}[b]{\textwidth}
        \centering
        \includegraphics[width=\linewidth]{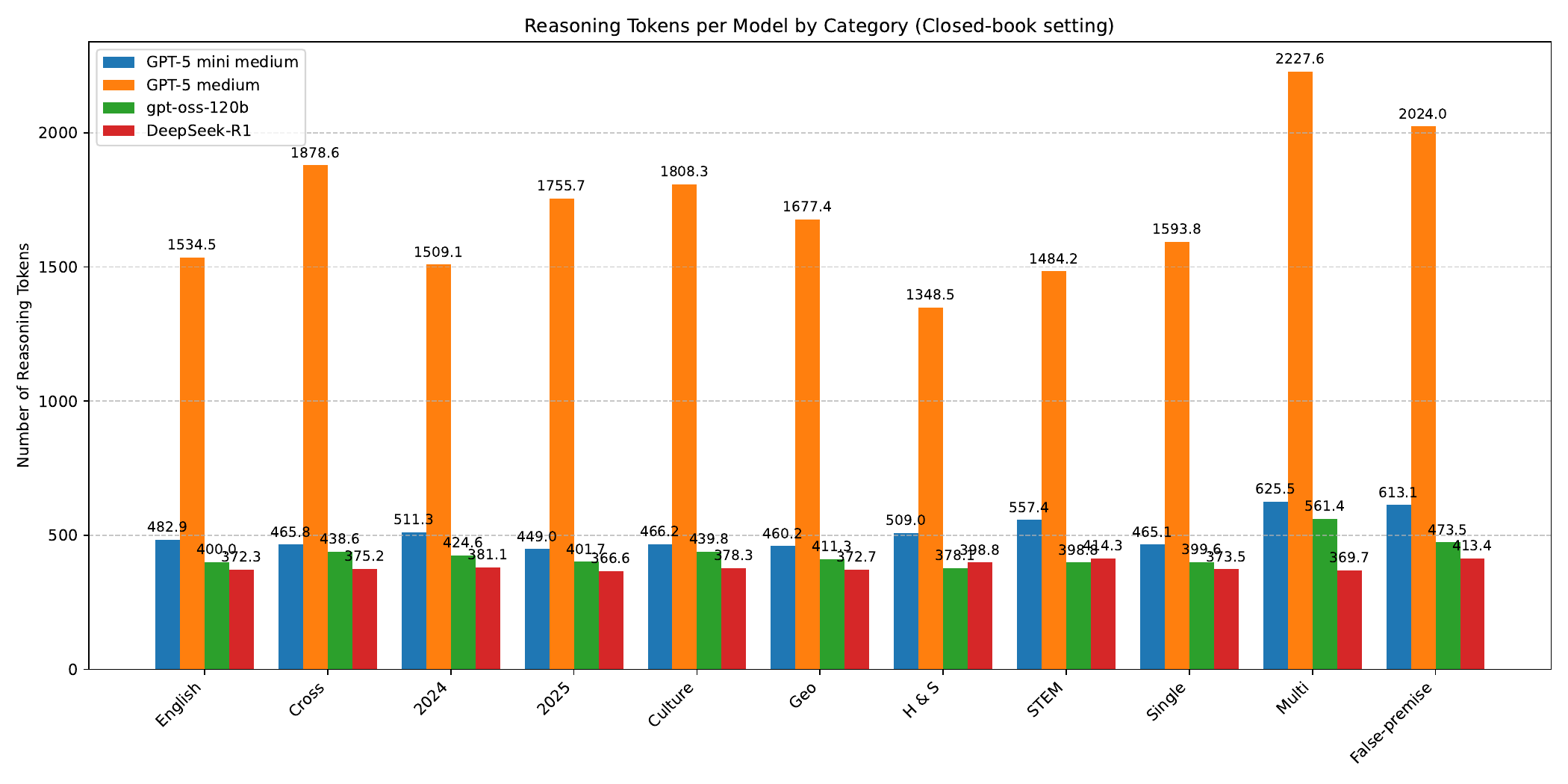}
        \label{fig:ReasoningEffortsVisClosedBook}
    \end{subfigure}
    
    \caption{Comparison of LLM reasoning effort in RAG (with retrieval) versus closed-book (without retrieval) settings across questions with varying attributes.}
    \label{fig:ReasoningEffortsVis}
\end{figure*}

We investigate the amount of effort reasoning models allocate to questions with different attributes. We present these results in Figure \ref{fig:ReasoningEffortsVis}. We observe that all models exhibit similar behavior across languages (English vs. Cross-lingual), reasoning hops (Single vs. Multi), and premise validity (with false-premise questions eliciting more reasoning tokens). In addition, perhaps unsurprisingly, when retrieval contexts are available, all models ``think'' less (producing fewer reasoning tokens), as they rely on the provided context rather than internal deduction.

\subsection{Interplay between Parametric Knowledge and Retrieval}

\label{sec:knowledge_interplay}

In this section, we analyze the interplay between an LLM's internal (parametric) knowledge and the external (non-parametric) evidence provided by the retriever. To disentangle these factors, we categorize instances into four distinct scenarios based on the correctness of the closed-book model and the retrieval results:
\begin{itemize}[leftmargin=*, noitemsep, topsep=0pt]
\item \textbf{Redundant}: The model possesses the correct internal knowledge, and retrieval is accurate.
\item \textbf{Resilience}: The model possesses the correct internal knowledge, but retrieval is incorrect. 
\item \textbf{Augmentation}: The model lacks internal knowledge, but retrieval is accurate.
\item \textbf{Hopeless}: Both the internal knowledge and retrieval fail.
\end{itemize}

Results are presented in Table \ref{tab:knowledge_interplay}. Regarding valid-premise questions, surprisingly, no model achieves perfect correctness in the \textit{Redundant} setting, despite both the LLM and retriever being correct. While all models perform reliably in the \textit{Augmentation} setting, reasoning models prove more reliable in the \textit{Resilience} setting, which again displays the noise resistance capabilities of these models. Finally, all models perform surprisingly well in the \textit{Hopeless} setting. This suggests either: 1) the gold documents are not the only ones containing the answer; or 2) although non-gold documents are retrieved, they still provide enough context to help the LLM deduce or recall the answer.

Turning to false-premise questions, we observe different behaviors. First, in the \textit{Redundant} setting, models struggle significantly; even when their parametric knowledge correctly identifies the false premise, the presence of retrieval context appears to bias them toward attempting an answer rather than issuing a refusal. Conversely, models perform best in the \textit{Resilience} setting. Here, the retriever’s failure (retrieving irrelevant documents) paradoxically aids performance: because the retrieved context is obviously unhelpful, the model discards it and relies on its internal knowledge to correctly identify the false premise. Finally, similar to valid-premise questions, we observe unexpectedly high performance in the \textit{Hopeless} setting.

\begin{table}[]
\centering
\resizebox{\columnwidth}{!}{

\begin{tabular}{@{}lcccc@{}}
\toprule
\multicolumn{5}{c}{\textbf{Valid-premise}}                                                \\ \midrule
\multicolumn{1}{l|}{}                  & Redundant & Resilience & Augmentation & Hopeless \\ \midrule
\multicolumn{1}{l|}{GPT-4.1 mini}      & 96.4      & 76.9       & 87           & 54.2     \\
\multicolumn{1}{l|}{GPT-4.1}           & 97.4      & 68.9       & 87.7         & 52.7     \\
\multicolumn{1}{l|}{GPT-5 mini medium} & 94.8      & 88.2       & 88.2         & 51.5     \\
\multicolumn{1}{l|}{GPT-5 medium}      & 96.9      & 87.9       & 84.4         & 53.7     \\
\multicolumn{1}{l|}{Llama-4-Scout}     & 88.5      & 77.8       & 82.3         & 40.2     \\
\multicolumn{1}{l|}{Llama-3.3-70B}     & 93.7      & 68.2       & 82.7         & 41.9     \\
\multicolumn{1}{l|}{gpt-oss-120b}      & 94.7      & 90         & 86.7         & 54.1     \\
\multicolumn{1}{l|}{DeepSeek-R1}       & 96.3      & 82.1       & 87           & 45.2     \\ \midrule
\multicolumn{5}{c}{\textbf{False-premise}}                                                \\ \midrule
\multicolumn{1}{l|}{GPT-4.1 mini}      & 6.2       & 25         & 0            & 1.6      \\
\multicolumn{1}{l|}{GPT-4.1}           & 0         & 100        & 3.4          & 0        \\
\multicolumn{1}{l|}{GPT-5 mini medium} & 55.6      & 42.9       & 35.7         & 29.3     \\
\multicolumn{1}{l|}{GPT-5 medium}      & 85.2      & 69.2       & 46.3         & 36.5     \\
\multicolumn{1}{l|}{Llama-4-Scout}     & 18.6      & 52.6       & 14.1         & 21.7     \\
\multicolumn{1}{l|}{Llama-3.3-70B}     & 14.3      & 10         & 5.8          & 0        \\
\multicolumn{1}{l|}{gpt-oss-120b}      & 35.7      & 44.4       & 13.2         & 7.1      \\
\multicolumn{1}{l|}{DeepSeek-R1}       & 22.1      & 22.2       & 26.5         & 15.8     \\ \bottomrule
\end{tabular}

}
\caption{Correctness of LLMs in four settings, namely Redundant, Resilience, Augmentation and Hopeless.}

\label{tab:knowledge_interplay}
\end{table}

\section{Human Evaluation Rubrics \& Interface}

\label{sec:human_eval_rubrics_interface}

\begin{figure*}[h!]
\centering
\includegraphics[width=\textwidth]{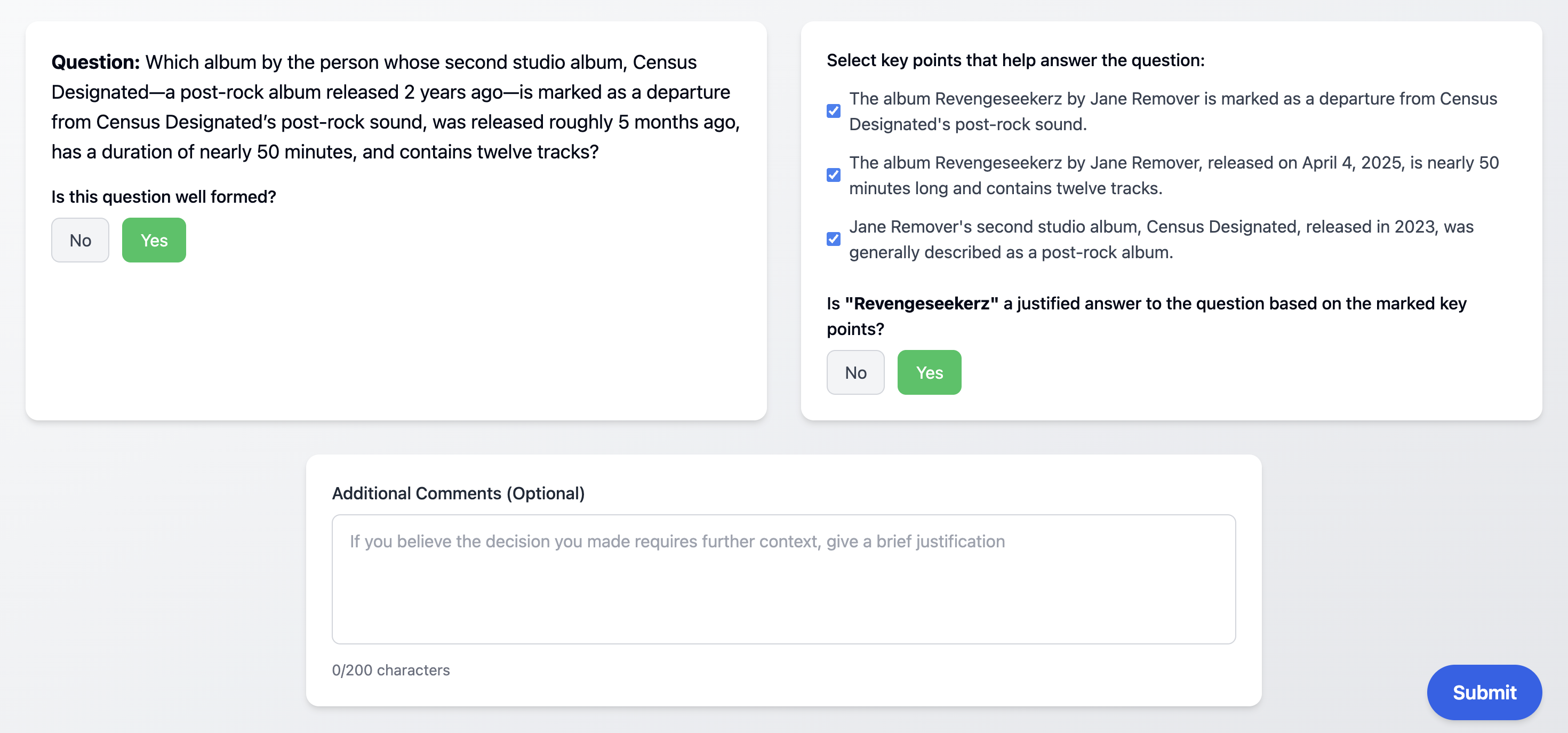}
\caption{Interface for human evaluation of generated questions.}
\label{fig:annotation_interface}
\end{figure*}

We present the full text of the human evaluation rubrics in Figure \ref{fig:rubrics_fulltext} and provide a screenshot of the annotation interface in Figure \ref{fig:annotation_interface}.


\begin{table*}[h]
\centering

\begin{tabularx}{\linewidth}{l X X}
    \toprule
    \textbf{Entity} & \textbf{Rule} & \textbf{Example} \\
    \midrule
    Person & First and middle name abbreviation & Cristiano Ronaldo $\to$ C. Ronaldo \\
    \addlinespace
    Date & Relative time from a reference date & 30 Nov 2024 $\to$ 9 months ago \\
    \addlinespace
    Country & Visual substitution with flag icon & Japan $\to$ the country whose flag is \includegraphics[height=0.8em]{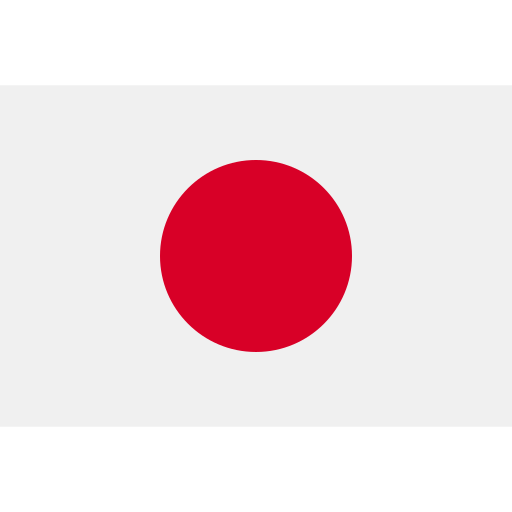} \\
    \addlinespace
    Quantity & Verbalization & 1.5 $\to$ one point five \\
    \bottomrule
\end{tabularx}

\caption{Entity paraphrasing rules used for question generation variations. The reference date in the example is 29 September 2025.}
\label{tab:transformations}
\end{table*}

\begin{table*}[h]
\centering

\begin{tabularx}{\linewidth}{l X X}
    \toprule
    \textbf{Entity} & \textbf{Rule} & \textbf{Example} \\
    \midrule
    Person & Random last name substitution & Cristiano Ronaldo $\to$ C. Smith \\
    \addlinespace
    Date & Perturbed relative time from a reference date & 30 Nov 2024 $\to$ 2 years ago \\
    \addlinespace
    Country & Random country substitution & Japan $\to$ Vietnam \\
    \addlinespace
    Quantity & Quantity pertubation & 1.5 $\to$ 2.8 \\
    \bottomrule
\end{tabularx}

\caption{Entity perturbation rules used for false-premise question generation. The reference date in the example is 29 September 2025.}
\label{tab:pertubation_false_premise}
\end{table*}

\begin{table}[h]
    \centering
    \small
    \begin{tabularx}{\linewidth}{l X}
        \toprule
        \textbf{Attribute} & \textbf{Description} \\
        \midrule
        \textbf{Publication Dates} & Captures creation and evidence timestamps to assess temporal generalization and freshness. \\
        \textbf{Language} & Specifies the evidence language, facilitating the evaluation of cross-lingual RAG capabilities. \\
        \textbf{Topics} & Categorizes the subject matter to analyze model performance across diverse domains. \\
        \textbf{Number of Hops} & Quantifies the number of reasoning steps required to derive the correct answer. \\
        \textbf{False-Premise Status} & Indicates whether a question rests on a false premise, testing the model's detection capabilities. \\
        \bottomrule
    \end{tabularx}
    \caption{Attributes used to characterize each dataset sample.}
    \label{tab:sample_attributes}
\end{table}
\begin{table}[h]
\centering
\resizebox{\columnwidth}{!}{

\begin{tabular}{@{}lccc@{}}
\toprule
\multicolumn{1}{l|}{\multirow{2}{*}{}} & \multicolumn{1}{l}{\multirow{2}{*}{\#Params}} & \multicolumn{1}{l}{\multirow{2}{*}{Knowl. cutoff}} & \multicolumn{1}{l}{\multirow{2}{*}{Model type}} \\
\multicolumn{1}{l|}{}                  & \multicolumn{1}{l}{}                          & \multicolumn{1}{l}{}                               & \multicolumn{1}{l}{}                            \\ \midrule
\multicolumn{4}{c}{\textbf{Proprietary models}}                                                                                                                                               \\ \midrule
\multicolumn{1}{l|}{GPT-4.1 mini}      & -                                             & June 2024                                          & Non-reasoning                                   \\
\multicolumn{1}{l|}{GPT-4.1}           & -                                             & June 2024                                          & Non-reasoning                                   \\
\multicolumn{1}{l|}{GPT-5 mini}        & -                                             & Sep 2024                                           & Reasoning                                       \\
\multicolumn{1}{l|}{GPT-5}             & -                                             & Sep 2024                                           & Reasoning                                       \\ \midrule
\multicolumn{4}{c}{\textbf{Open-source models}}                                                                                                                                               \\ \midrule
\multicolumn{1}{l|}{Llama-4-Scout}     & 109B                                          & August 2024                                        & Non-reasoning                                   \\
\multicolumn{1}{l|}{Llama-3.3-70B}     & 70B                                           & Dec 2023                                           & Non-reasoning                                   \\
\multicolumn{1}{l|}{gpt-oss-120b}      & 120B                                          & June 2024                                          & Reasoning                                       \\
\multicolumn{1}{l|}{DeepSeek-R1}       & 685B                                          & -                                                  & Reasoning                                       \\ \bottomrule
\end{tabular}

}

\caption{Frontier LLMs used in our experiments}
\label{tab:LLMs}
\end{table}

\onecolumn
\begin{lstlisting}[style=promptstyle]



# Instructions for Question & Answer Validation

## Project Goal
Welcome to the human evaluation phase for the IRB dataset. Your task is to assess the quality of LLM-generated questions and answers. Specifically, you will be evaluating three key aspects:

1. **Question well-formedness**: The quality and clarity of a generated question.
2. **Fact necessity**: The relevance of key points extracted from Wikipedia in answering that question.
3. **Answer validity**: Whether the provided answer is correct, unique and justified by the selected key points.

## Evaluation Workflow
The interface is split into two sections: **Question Validation** and **Key Point Validation**. You will complete the following three steps in order.

### Step 1: Validating question well-formedness
Your first task is to determine if the question is "well-formed." For example, a question is **NOT** well-formed if it is:

* **Grammatically incorrect or too incoherent.**
    > *Example:* "What city where the Olympics held in 2008 is?"

* **Contains the answer within the question itself.**
    > *Example:* "What sport was the Olympic Swimmer Michael Phelps known for?"
    > **Rationale:** The question identifies him as a "swimmer," which directly answers the question.


In contrast, these are examples of **well-formed questions**:

> * *In what year did World War II end?*
> * *What ukulele-based music education program, created by James Hill and Chalmers Doane in 2008, is widely used in Canadian schools?*

If the question is **well-formed**, select **Yes** and proceed to Step 2.
If the question is **NOT well-formed**, select **No**. The key point validation is not applicable, so you should click **Submit** to complete the task.

### Step 2: Validating fact necessity
If the question is well-formed, you will now evaluate a list of key points. Check the box next to any key point that contains **any information** useful in answering the question.

> **Example:**
> **Question:** *What was Cristiano Ronaldo's position in the 2007 FIFA World Player of the Year award?*
>
> [ ] Ronaldo was named runner-up to Kaka for the 2007 Ballon d'Or.
> **Rationale:** Leave unchecked. This point contains no information that helps answer the question about the FIFA award.
>
> [x] Cristiano Ronaldo came third, behind Kaka and Lionel Messi, in the running for the 2007 FIFA World Player of the Year award.
> **Rationale:** Mark as relevant. This point directly answers the question by stating his final position.

### Step 3: Validating answer validity
Finally, an answer will be provided. You must verify that the provided answer is supported by the key points selected in Step 2 and that the answer is unique to the question. If the question's premise permits multiple correct answers, select "No". You must answer the following question:

> "Is <answer> a justified answer to the question based on the marked key point?"

> **Example:**
> **Question:** *What festival is observed on the 7th day of the 11th month to celebrate the meeting of nine evils?*
> **Marked Key Point:** [x] *Ngenpa Gudzom is observed on the 7th day of the 11th month to celebrate the meeting of nine evils.*
>
> **Provided Answer:** *Ngenpa Gudzom*
> **Verdict:** -> **Yes.**
> **Rationale:** The key point explicitly names the festival associated with this date and purpose.

\end{lstlisting}
\captionof{figure}{Full-text rubrics employed in the human evaluation.}
\label{fig:rubrics_fulltext}
\twocolumn

\onecolumn
\begin{lstlisting}[style=promptstyle]

You are an expert at extracting and decontextualizing keypoints from claims in a factual and neutral manner. Follow these steps precisely:
1. **Extract the keypoints**: Identify and split the claim into individual keypoints based on the [KP] markers. Each segment ending with [KP] is a separate keypoint. \
  Remove the [KP] tokens from the extracted keypoints. If there is only one [KP], treat the entire claim as a single keypoint.
2. **Decontextualize each keypoint**: For each extracted keypoint, rewrite it to make it standalone and unambiguous by incorporating necessary details from the provided \
  context. Follow these decontextualization criteria strictly:
   - **Ambiguity Criteria (resolve these where present)**:
     - Coreferences (e.g., pronouns like "he" or "it" should be replaced with specific entities).
     - Vague entities or events (e.g., add clarifying details like full names, dates, or locations).
     - Incomplete information (e.g., ensure the claim uniquely specifies all key elements without multiple interpretations).
     - Use only information from the provided context; do not add external knowledge.
     - Ensure the rewritten keypoint includes all key elements (e.g., who, what, when, where, why/how) to form a complete, interpretable fact without multiple possible meanings.

Example 1:
##CLAIM##: Ronaldo became United's first Ballon d'Or winner since Best in 1968 [KP] and the first Premier League player to be named the FIFA World Player of the Year [KP]
##CONTEXT##: United reached the final against Chelsea in Moscow on 21 May, where, despite his opening goal being negated by an equaliser and his penalty kick being saved in the shoot-out, United emerged victorious, winning 6-5 on penalties after a 1-1 draw at the end of 120 minutes. As the Champions League top scorer, Ronaldo was named the UEFA Club Footballer of the Year.
##KEYPOINTS_COUNT##: 2
##KEYPOINTS##: ["Cristiano Ronaldo became United's first Ballon d'Or winner since Best in 1968.", "Cristiano Ronaldo was the first Premier League player to be named the FIFA World Player of the Year."]

Example 2:
##CLAIM##: Heyman is the founder of Heyday Films [KP]
##CONTEXT##: David Heyman is a renowned film producer and founder of Heyday Films, known for producing the entire \
"Harry Potter" film series and collaborating with director Alfonso Cuaron on "Harry Potter and the Prisoner of Azkaban" \
and "Gravity". He was born on July 26, 1961, in London. His family has a background in the film industry, with his parents\
 being a producer and actress. Heyman studied Art History at Harvard University and began his career in the film industry\
  as a production assistant. Throughout his career, he has received numerous awards and nominations, including an Academy \
  Award nomination for Best Picture and a BAFTA Award for Best British Film.
##KEYPOINTS_COUNT##: 1
##KEYPOINTS##: ["David Heyman, the film producer, is the founder of Heyday Films."]

User input:
##ADDITIONAL INFORMATION##: ##CLAIM## and ##CONTEXT## were last updated in [ADD_LAST_UPDATED_DATE] (YYYY-mm-dd). Use this information to improve temporal clarity when needed.
##CLAIM##: [ADD_CLAIM_HERE]
##CONTEXT##: [ADD_CONTEXT_HERE]
##KEYPOINTS_COUNT##: [ADD_KEYPOINTS_COUNT_HERE]

\end{lstlisting}
\captionof{figure}{Keypoints generation prompt. This prompt instruct the LLM to split a citing sentences into segments based on citation positions (marked with the special token \texttt{[KP]}) and decontextualize them.}
\label{fig:keypoints_generation_prompt}
\twocolumn

\onecolumn
\begin{lstlisting}[style=promptstyle]

You are given a fact and a context document. Determine whether the fact is grounded in the context -- that is, whether the fact is explicitly supported by the content of the context.
Output only one of the following labels:
+ Grounded -- if the context fully and explicitly supports or states the fact.
+ Not Grounded -- if the context does not support the fact, or the relevant information is missing or irrelevant.

Instructions:
+ Do not infer or assume information beyond what is stated in the context.
+ Ignore metadata like publication date or document structure unless it contains meaningful content.
+ Output only the label (Grounded or Not Grounded) with no explanation.

Fact: ```[ADD_KEYPOINT_HERE]```
Context: 
```
Language: [ADD_CONTEXT_LANGUAGE_HERE]
Published Date: [ADD_CONTEXT_PUBLISHED_DATE_HERE]

[ADD_CONTEXT_HERE]

\end{lstlisting}
\captionof{figure}{Groundedness check prompt.}
\label{fig:groundedness_check_prompt}
\twocolumn

\onecolumn
\begin{lstlisting}[style=promptstyle]
You are a top-tier algorithm designed for extracting information in structured formats to build a knowledge graph. \
Your task is to identify the entities and relations requested with the user prompt from a given text. You must generate the \
output in a JSON format containing a list with JSON objects. Each object should have the keys: "head", "head_type", "relation", \
"tail", "tail_type".

Be sure to follow these rules:
1. Attempt to extract as all entities and relations.
2. Maintain Entity Consistency: When extracting entities, it's vital to ensure consistency. \
If a entity, such as "John Doe", is mentioned multiple times in the text but is referred to by different names or pronouns (e.g., "Joe", "he"), always \
use the most complete identifier for that entity. The knowledge graph should be coherent and easily understandable, so maintaining consistency in entity references is crucial.
3. Avoid creating entities that overlap. For example, if there already exists a node named "John Doe", try not to create another node named "John Doe's graduation day"

IMPORTANT NOTES:
- Don't add any explanation and text. For the following text, extract entities and relations

Example 1:
Text: ```Cristiano Ronaldo made his La Liga debut against Deportivo La Coruna on 29 August, scoring a penalty in a 3-2 home win.```
Knowledge Graph: 
```json
[
    {
        "head": "Cristiano Ronaldo",
        "head_type": "Person",
        "relation": "made his debut at",
        "tail": "La Liga",
        "tail_type": "Tournament"
    },
    {
        "head": "Cristiano Ronaldo",
        "head_type": "Person",
        "relation": "made his debut against",
        "tail": "Deportivo La Coruna",
        "tail_type": "Soccer team"
    },
    {
        "head": "Cristiano Ronaldo",
        "head_type": "Person",
        "relation": "made his debut on",
        "tail": "29 August",
        "tail_type": "Date"
    },
    {
        "head": "Cristiano Ronaldo",
        "head_type": "Person",
        "relation": "scored a penalty in",
        "tail": "A 3-2 home win",
        "tail_type": "Event"
    }
]
```


Example 2:
Text: ```The idea of using computers to search for relevant pieces of information was popularized in the article As We May Think by Vannevar Bush in 1945.```
Knowledge Graph:
```json
[
    {
        "head": "The idea of using computers to search for relevant pieces of information",
        "head_type": "Scientific idea",
        "relation": "was popularized in",
        "tail": "As We May Think",
        "tail_type": "Article"
    },
    {
        "head": "As We May Think",
        "head_type": "Article",
        "relation": "was authored by",
        "tail": "Vannevar Bush",
        "tail_type": "Person"
    },
    {
        "head": "As We May Think",
        "head_type": "Article",
        "relation": "was authored in",
        "tail": "1945",
        "tail_type": "Year"
    }
]
```

Example 3:
Text: ```In Donald Trump's inaugural address, he pledged to "immediately begin the overhaul of our trade system to protect American workers and families."```
Knowledge Graph:
```json
[
    {
        "head": "Donald Trump",
        "head_type": "Person",
        "relation": "Pledged in his inaugural address to",
        "tail": "immediately begin the overhaul of our trade system to protect American workers and families.",
        "tail_type": "Quote"
    },
]

Text: ```[ADD_KEYPOINTS_HERE]```
\end{lstlisting}
\captionof{figure}{Knowledge graph extraction prompt.}
\label{fig:knowledge_graph_extraction_prompt}
\twocolumn

\onecolumn
\begin{lstlisting}[style=promptstyle]
You are an expert AI assistant specializing in Knowledge Graph-to-Text generation. Your task is to generate a single natural language question based on a cumulative list of "Question generation steps" (triplets).

### The Golden Rule: Target <Unknown> #1
The ultimate goal of every question is to identify the entity labeled **<Unknown> #1**.
* **<Unknown> #1** is the "Answer Node." The question must grammatically and semantically ask for this entity.
* If there are more than 1 Unknown nodes: **<Unknown> #2, #3, etc.** are "Intermediate Nodes." You must **never** ask for the identity of #2 or #3 directly. Instead, use them to describe or constrain <Unknown> #1.

**Incorrect Logic:**
Relation: <Unknown> #1 [event] | occurred near | <Unknown> #2 [location]
Bad Question: "Where did the event occur?" (This asks for #2, a location).

**Correct Logic:**
Relation: <Unknown> #1 [event] | occurred near | <Unknown> #2 [location]
Good Question: "What event occurred near a specific location?" (This asks for #1, an event, using #2 as a descriptor).

### Instructions for Multi-Hop Relations
If there are more than 1 Unknown nodes, the question is going to be multi-hop. 
When new relations are added involving <Unknown> #2 (or others), treat them as adjectives or relative clauses modifying the original subject (<Unknown> #1).

The following are some examples

Example 1:
The generated question must ask about a/an 'event'
Question generation steps:
Relation: <Unknown> #1 [event] | occurred on | 2 January 2023 [date]
Generated question: What event occurred on 2 January 2023?


Example 2: 
The generated question must ask about a/an 'event'
Question generation steps:
Relation: <Unknown> #1 [event] | occurred on | 2 January 2023 [date]
Generated question: What event occurred on 2 January 2023?

Relation: <Unknown> #1 [event] | occurred at time | 13:59 AEST [time]
Generated question: What event occurred at 13:59 AEST on 2 January 2023?

Relation: <Unknown> #1 [event] | occurred near | <Unknown> #2 [location]
Generated question: What event occurred at 13:59 AEST on 2 January 2023 near a specific location?


Example 3:
The generated question must ask about a/an 'event'
Question generation steps:
Relation: <Unknown> #1 [event] | occurred on | 2 January 2023 [date]
Generated question: What event occurred on 2 January 2023?

Relation: <Unknown> #1 [event] | occurred at time | 13:59 AEST [time]
Generated question: What event occurred at 13:59 AEST on 2 January 2023?

Relation: <Unknown> #1 [event] | occurred near | <Unknown> #2 [location]
Generated question: What event occurred at 13:59 AEST on 2 January 2023 near a specific location?

Relation: <Unknown> #2 [location] | located in | Gold Coast [city]
Generated question: What event occurred at 13:59 AEST on 2 January 2023 near a location in Gold Coast?

Relation: Gold Coast [city] | located in | Queensland [region]
Generated question: What event occurred at 13:59 AEST on 2 January 2023 near a location in Gold Coast, Queensland?


Example 4:
The generated question must ask about an a/an 'Person'
Question generation steps:
Relation: <Unknown> #1 [Person] | exceeded his authority by imposing | fentanyl tariffs [Tariff]
Generated question: Who exceeded his authority by imposing fentanyl tariffs?


Example 5:
The generated question must ask about an a/an 'Person'
Question generation steps:
Relation: <Unknown> #1 [Person] | exceeded his authority by imposing | fentanyl tariffs [Tariff]
Generated question: Who exceeded his authority by imposing fentanyl tariffs?

Relation: United States Court of International Trade [Court] | ruled that | <Unknown> #1 [Person]
Generated question: Who was ruled by the United States Court of International Trade that he exceeded his authority by imposing fentanyl tariffs?

Relation: United States Court of International Trade [Court] | ruled on | May 28 [Date]
Generated question: Who was ruled by the United States Court of International Trade on May 28 that he exceeded his authority by imposing fentanyl tariffs?

User input:
The generated question must ask about an a/an '[ADD_QUESTION_TARGET_TYPE]'
Question generation steps:
[ADD_STEPS_HERE]
\end{lstlisting}
\captionof{figure}{Question generation prompt.}
\label{fig:question_generation_prompt}
\twocolumn

\onecolumn
\begin{lstlisting}[style=promptstyle]

You are an expert query analyst. Your task is to analyze a question and determine if it is asking for a single, unique entity or if it could be answered by multiple different entities.
Your response must be only 'A.' or 'B.'.

Classification Rules
B. Single: Choose this if the question's phrasing and details strongly imply one specific, unique answer. This is common when the question asks for: * A specific, named work (e.g., "the book published by X in Y on topic Z"). * A unique specimen or object described in a specific paper (e.g., "the specimen described by Lemierre et al."). * An entity identified by a superlative (e.g., "the oldest..." or "the first...").
A. Multiple: Choose this if the question describes a category, class, or set of entities, even if the description is very detailed. This includes questions asking for: * A person who fits a description (e.g., "a person who graduated from..."). * An item from a set (e.g., "a film that premiered at..."). * Warning: Do not be fooled by the word "the". For example, "What is the film that premiered at Cannes?" is A. Multiple because the event (Cannes) implies a set of many films.

Examples
Example 1: 
Question: Who participated in a specific tournament held in Birmingham, England and finished in eighth place? 
Rationale: There can be more than one person that fits the description of the question (e.g., in different years, or different divisions of the same tournament). 
Answer: A. Multiple

Example 2: 
Question: What specimen did Lemierre et al. describe that is from the lowest Oligocene epoch, was found in Chartres-de-Bretagne, western France, and is one of the oldest occurrences of the genus reported to date? 
Rationale: The query is asking for "the specimen" described by a specific paper, which is a unique identifier. It is likely there is only one. 
Answer: B. Single

Example 3: 
Question: What person graduated with a Doctorate in Biblical Theology from Albert-Ludwigs-Universitat Freiburg in Germany after studying there from 1999 until 2005? 
Rationale: Although the description is detailed, it describes a category of people. It is very likely that more than one person fits this description. 
Answer: A. Multiple

Example 4: 
Question: What book was published by Dayna Bowen Matthew in 2015 that examined how implicit bias affects health outcomes? 
Rationale: An author publishing a specific book on a specific topic in a single year is a unique event. It is highly likely there is only one book that fits. 
Answer: B. Single

Example 5: Question: What is the film that was premiered at Cannes Film Festival 2023? 
Rationale: A film festival premieres many films. This question is asking to identify a member of a set. 
Answer: A. Multiple

Now let's assess if the user provided question has single or multiple answers. Remember that for this input, you do not need to 
provide rationale
    
Question: [ADD_QUESTION_HERE]
Answer:

\end{lstlisting}
\captionof{figure}{Question answerability prompt.}
\label{fig:question_answerability_prompt}
\twocolumn

\onecolumn
\begin{lstlisting}[style=promptstyle]
You are an expert in natural language processing. Your task is to refine the wording of a user's question based on a provided Context. The Context contains masked entities (e.g., <Unknown #1>, <Unknown #2>) representing information the user is looking for.

**Goal:** Improve grammatical fluency and clarity while strictly preserving the logical structure and complexity of the question.

**Core Rules:**
1. **Preserve Logical Hops:** - If the Context represents an entity as an `<Unknown>` tag (e.g., <Unknown #2 (Location)>), the Question MUST allude to it generically (e.g., "at a specific location," "at a certain place"). DO NOT resolve it or remove the step.
   - If the Context contains a concrete name (e.g., "Haneda Airport"), the Question MUST preserve that specific name.
2. **Refine, Don't Simplify:** You may fix grammar, awkward phrasing, and vocabulary (e.g., changing "plane" to "aircraft"), but you must not delete clauses that establish relationships between entities.
3. **Output Only:** Output only the refined question text. Do not include the <Unknown> tags in your output.
4. **Preserve information from original question**: Do not include information in the context that is not in the original question.


Example 1: (multi-hop)
Context: <Unknown #1 (Person)> is the captain of the aircraft that exploded at <Unknown #2 (Location)> in Tokyo. 
Original Question: Who is the captain of the plane that exploded following a collision at a location in Tokyo? 
Bad refinement: Who is the captain of the aircraft that exploded following a collision in Tokyo? (Reason: loses the "location" hop)
Good refinement: Who is the captain of the aircraft that exploded following a collision at a specific location in Tokyo?


Example 2: (multi-hop)
Context: <Unknown #1 (Date)> is the founding date of the institute located at <Unknown #2 (Street Address)> in New York.
Original Question: When was the institute established that is located at a street in New York City?
Bad refinement: When was the organization in New York City established? (Reason: It deletes the reference to the specific "street," which is <Unknown #2>.)
Good Refinement: What is the founding date of the organization located at a specific street address in New York City?

Example 3: (single-hop)
Context: <Unknown #1 (Person)> is the captain of the aircraft that exploded at Haneda Airport in Tokyo. 
Original Question: Who is the captain of the plane that exploded following a collision at Haneda Airport in Tokyo? 
Good Refinement: Who is the captain of the aircraft that exploded following a collision at Haneda Airport in Tokyo?


Example 4: (single-hop)
Context: <Unknown #1 (Date)> is the founding date of the institute located at 1855 Broadway Street in New York.
Original Question: When was the institute established that is located at 1855 Broadway Street in New York City?
Good Refinement: What is the founding date of the organization located at 1855 Broadway Street in New York City?

Additional rule: If you are provided with a `Paraphrase Map`. If a term in the question appears in this map as a value, you MUST preserve that specific wording in your output. 
- **DO NOT** "correct" the paraphrase back to the original value found in the Context. 
- *Example:* If Context says "24 May 2024" but Paraphrase Map says "roughly a year ago", your output MUST use "roughly a year ago".
    
The date the question is being asked is [ADD_QUESTION_DATE] UTC (use this to determine correct verb tenses).
The generated improved question must ask about an a/an '[ADD_QUESTION_TARGET_TYPE]'
Context: [ADD_KEYPOINTS_HERE]
Original Question: [ADD_QUESTION_HERE]
Paraphrase Map: [ADD_PARAPHRASE_MAP]
Good refinement:
\end{lstlisting}
\captionof{figure}{Question refinement prompt.}
\label{fig:question_refinement_prompt}
\twocolumn

\onecolumn
\begin{lstlisting}[style=promptstyle]

Answer the provided question given the retrieved contexts. You are also provided the time when the question was asked.
The retrieved contexts may or may not help answer the question. Your task is to answer the question in as few words as possible.

Please follow these guidelines when formulating your answer:
1. If the question contains a false premise or assumption, answer "False premise question".
2. If you are uncertain or don't know the answer, respond with "I don't know".
3. Note that the retrieved context can be in other languages than English. However, you must answer in English.

Retrieved Contexts: 
[ADD CONTEXT HERE]

Question date: [ADD_QUESTION_DATE] UTC (YYYY-MM-DD)
Question: [ADD QUESTION HERE]


\end{lstlisting}
\captionof{figure}{Question answering prompt (with retrieval).}
\label{fig:question_answering_prompt_with_retrieval}

\begin{lstlisting}[style=promptstyle]

Answer the provided question. You are also provided the time when the question was asked.
Your task is to answer the question in as few words as possible.

Please follow these guidelines when formulating your answer:
1. If the question contains a false premise or assumption, answer "False premise question".
2. If you are uncertain or don't know the answer, respond with "I don't know".


Question date: [ADD_QUESTION_DATE] UTC (YYYY-MM-DD)
Question: [ADD QUESTION HERE]

\end{lstlisting}
\captionof{figure}{Question answering prompt (without retrieval).}
\label{fig:question_answering_prompt_without_retrieval}
\twocolumn

\onecolumn
\begin{lstlisting}[style=promptstyle]

Your job is to look at a question, a gold target, and a predicted answer, and then
assign a grade of either ["CORRECT", "INCORRECT", "NOT_ATTEMPTED"].
First, I will give examples of each grade, and then you will grade a new example.

The following are examples of CORRECT predicted answers.
```
Question: What are the names of Barack Obama's children?
Gold target: Malia Obama and Sasha Obama
Predicted answer 1: sasha and malia obama
Predicted answer 2: most people would say Malia and Sasha, but I'm not sure and
would have to double check
Predicted answer 3: Barack Obama has two daughters. Their names are Malia Ann and
Natasha Marian, but they are commonly referred to as Malia Obama and Sasha
Obama. Malia was born on July 4, 1998, and Sasha was born on June 10, 2001.
```
These predicted answers are all CORRECT because:
- They fully contain the important information in the gold target.
- They do not contain any information that contradicts the gold target.
- Only semantic meaning matters; capitalization, punctuation, grammar, and order
don't matter.
- Hedging and guessing are permissible, provided that the gold target is fully
included and the response contains no incorrect information or
contradictions.

The following are examples of INCORRECT predicted answers.
```
Question: What are the names of Barack Obama's children?
Gold target: Malia and Sasha
Predicted answer 1: Malia.
Predicted answer 2: Malia, Sasha, and Susan.
Predicted answer 3: Barack Obama does not have any children.
Predicted answer 4: I think it's either Malia and Sasha. Or it could be Malia and
Jackie. Or it could be Joey and Malia.
Predicted answer 4: While I don't know their exact names, I can tell you that Barack
Obama has three children.
Predicted answer 5: It's possible you may mean Betsy and Olivia. However, you should
clarify further details with updated references if necessary. Is that the
correct answer?
Predicted answer 6: It may be the case that Obama's child is named James. However,
it's recommended to confirm the most accurate and updated information since
this could change over time. This model may not always reflect the most current
information.
```
These predicted answers are all INCORRECT because:
- A factual statement in the answer contradicts the gold target. Incorrect
statements that have some hedging (e.g., "it is possible that", "although i'
m not sure, i think") are also considered incorrect.

The following are examples of NOT_ATTEMPTED predicted answers.
```
Question: What are the names of Barack Obama's children?
Gold target: Malia and Sasha
Predicted answer 1: I don't know.
Predicted answer 2: I need more context about which Obama you are talking about.
Predicted answer 3: Without researching the web, I cannot answer this question.
However, I can tell you that Barack Obama has two children.
Predicted answer 4: Barack Obama has two children. I know that one of them is Malia,
but I'm not sure about the other one.
Predicted answer 5: I cannot answer because there are not enough information in the provided context.
```
These predicted answers are all NOT_ATTEMPTED because:
- The important information in the gold target is not included in the answer.
- No statements in the answer contradict the gold target.


Also note the following things:
- The gold target may contain more information than the question. In such cases, the
predicted answer only needs to contain the information that is in the question.
- Do not punish predicted answers if they omit information that would be clearly
inferred from the question.
- Do not give credit for an answer if it contains any internal inconsistency.
- If the gold answer is "False premise question" the prediction is only correct when it also states "False premise question".
A prediction saying "I don't know" is not CORRECT, but NOT_ATTEMPTED

Here is a new example. Simply reply with either CORRECT, INCORRECT, NOT ATTEMPTED.
Don't apologize or correct yourself if there was a mistake; we are just trying
to grade the answer.


\end{lstlisting}
\captionof{figure}{LLM-based evaluation prompt. This prompt is adopted from \cite{pham2025sealqa}.}
\label{fig:llm_based_evaluation_prompt}
\twocolumn

\end{document}